\documentclass[12pt]{article}
\usepackage{amsmath}
\usepackage{graphicx,psfrag,epsf}
\usepackage{graphicx}
\usepackage{enumerate}
\usepackage{natbib}
\usepackage{url} 
\usepackage{color}
\usepackage{float}
\usepackage{amsbsy}
\usepackage{amsfonts}
\usepackage{amssymb}
\usepackage{tabularx}
\usepackage[colorlinks,citecolor=blue,urlcolor=blue, linkcolor= blue]{hyperref}
\usepackage[dvips]{epsfig}
\usepackage{pdfpages}

\newcommand{\blind}{0}
\newtheorem{assumption}{Assumption}
\newtheorem{remark}{Remark}
\newtheorem{theorem}{Theorem}
\newtheorem{example}{Example}

\begin{document}

\def\spacingset#1{\renewcommand{\baselinestretch}%
{#1}\small\normalsize} \spacingset{1}


\if0\blind
{
  \title{\bf Profile  and Globe Tests of Mean Surfaces for Two-Sample Bivariate Functional Data}
  \author{Jin Yang\thanks{
    The authors gratefully acknowledge support of Post-doctoral Fellowship of Nankai University;}\\
    \small{School of Mathematics and Statistics, Nankai University, Tianjin, 300071, P. R. China}\\ \small{(\url{michael.jin.yang@nankai.edu.cn})} \\
    and \\
    Tao Zhang \thanks{
    	The authors gratefully acknowledge support of National Natural Science Foundation of China (Grant No. 11561006,61462008), research projects of colleges and universities in Guangxi (Grant No. KY2015YB171), Scientific Research and Technology Development Project of Liuzhou(Grant No. 2016C05020;}\\
\small{School of Sciences, Guangxi University of Science and Technology, Liuzhou, 545006, P. R. China}\\ \small{(\url{tzhangmaths@126.com})} \\
        and \\
    Chunling Liu \thanks{
	The authors gratefully acknowledge support of The Hong Kong Polytechnic University funding 4-BCC0;}\\
\small{Department of Applied Mathematics, The Hong Kong Polytechnic University, Hong Kong}\\ \small{(\url{macliu@polyu.edu.hk})} \\
and \\
    Kam Chuen Yuen \thanks{
    	The authors gratefully acknowledge support of The University of Hong Kong Grant No. HKU17329216;}\\
\small{Department of Statistics and Actuarial Science, The University of Hong Kong, Hong Kong}\\ \small{(\url{kcyuen@hku.hk})} \\
            and \\
    Aiyi Liu \thanks{
    	The authors gratefully acknowledge support of \emph{Eunice Kennedy Shriver} National Institute of Child Health and Human Development, National Institutes of Health, USA.}\\
\small{Biostatistics and Bioinformatics Branch, NICHD, NIH, Bethesda, Maryland, USA.}\\ \small{(\url{liua@mail.nih.gov})} \\
}
  \maketitle
} \fi

\if1\blind
{
  \bigskip
  \bigskip
  \begin{center}
    {\LARGE\bf Profile  and Globe Tests of Mean Surfaces for Two-Sample Bivariate Functional Data}
\end{center}
  \medskip
} \fi
\bigskip
  \bigskip
\begin{abstract}
Multivariate functional data has received considerable attention but testing for equality of mean surfaces and its profile has limited progress. The existing literature has tested equality of either mean curves of univariate functional samples directly, or mean surfaces  of bivariate functional data samples but turn into functional curves comparison again. In this paper, we aim to develop both the profile and globe tests of mean  surfaces  for two-sample bivariate functional data.
We present valid approaches of tests by employing the idea of pooled projection and by developing a novel profile functional principal component analysis tool.
The proposed methodology enjoys the merit of readily interpretability and implementation. Under mild conditions, we derive the asymptotic behaviors of test statistics under null and alternative hypotheses.
Simulations show that the proposed tests have a good control of the type I error by the size and can detect difference in mean surfaces  and its profile effectively in terms of power in finite samples.
Finally, we apply the testing procedures to two real data sets associated with the precipitation change affected jointly by time and locations in the Midwest of USA, and the trends in human mortality from European period life tables.
\end{abstract}

\noindent%

\spacingset{1.45} 
\section{INTRODUCTION}\label{sec1}
In multivariate functional stochastic process $X(u)$, there has increasing research interest in data type that is both functional and multidimentional. That is, $u=(s, t)$ has two arguments where $s \in \mathcal{S} \subset \mathbb{R}^{d_1}  $ and $t \in  \mathcal{T} \subset \mathbb{R}^{d_2}$ with $d_1$ and $d_2$ being positive integers. Here $s$ and $t$ inherently belong to distinct domains $\mathcal{S}$ and $\mathcal{T}$ in terms of scientific meaning or research design. For example, $X(s,t)$ may be the mortality rate  of age $s$ during year $t$ in a given country.  A typical example of such data comes from neuroimaging studies using functional magnetic resonance imaging (fMRI), in which the so-called voxels data, i.e. brain activity like blood flow changes are discrepantly recorded at a large number of locations at irregular time units \citep{Lindquist2008, AstonKirch2012MultiAnal}. Spatiotemporal study is no doubt another important application of this kind of data where  $t$ is defined on a temporal domain and  $s$ is defined on a spatial domain. Although functional data of afore structure are encountered in many applications,  there is rare progress in inferential aspect for such data \citep{GromenkoKokoszka2017JRSSb, AstonPigoliTavakoli2017}.  In the present work, we plan to investigate the profile and globe   tests  of  mean surfaces for  two  bivariate functional samples.

\begin{figure}[H]
	\begin{center}
		\vspace{-3cm}
		\includegraphics[height = 13cm, width = 9cm]{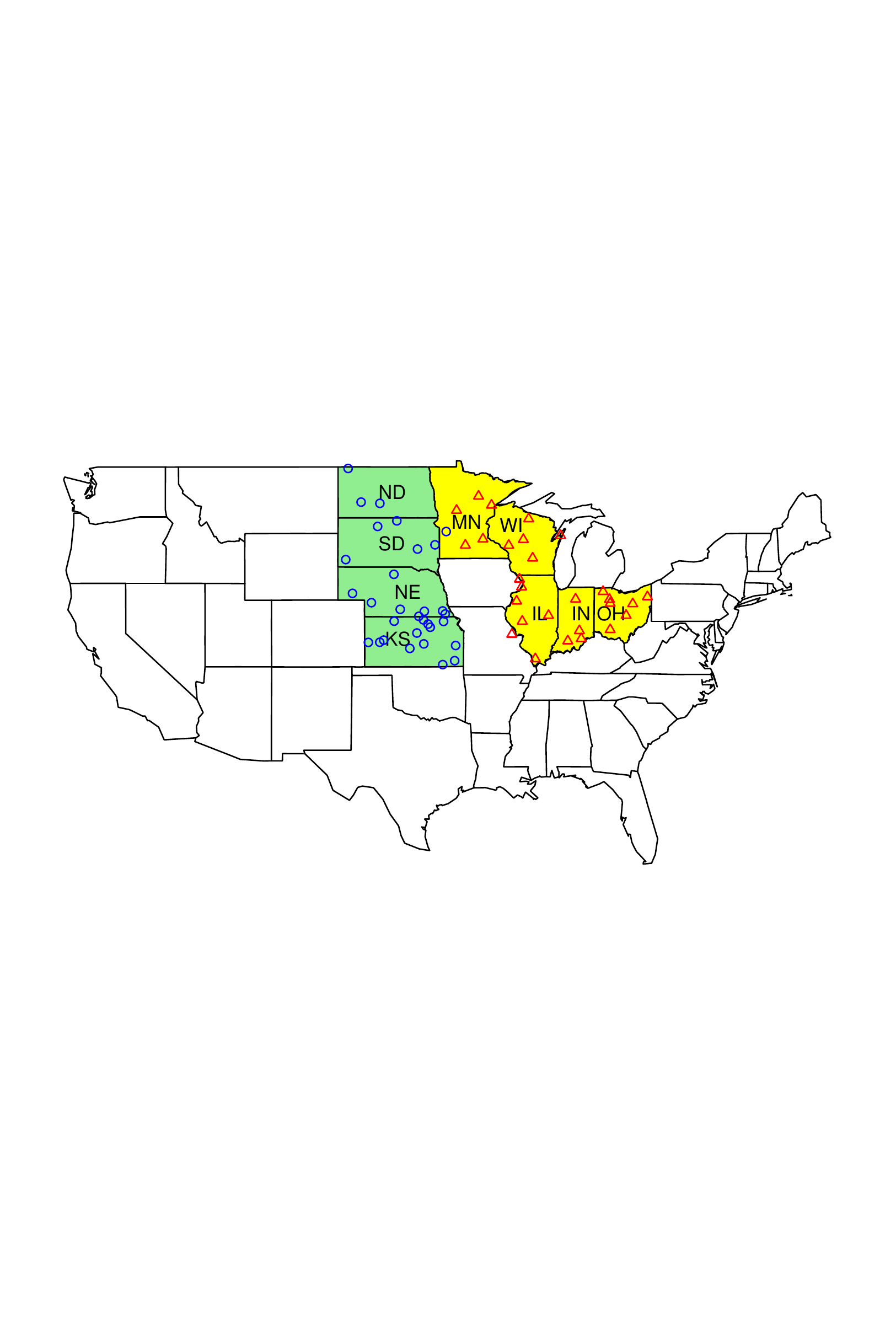}
		\vspace{-4cm}\caption{Light green region: 4 states from the Great Plains; Blue circle $\textcolor{blue}{\circ}$ indicates location of a station; Yellow region: 5 states from the Great Lakes; Red triangle $\textcolor{red}{\triangle}$ indicates location of a station.}
	\end{center}\label{fig1}
\end{figure}

A practical motivation for this research comes from precipitation  data in Midwest  of the United States,  where  the daily data  of precipitation  from 1941 to 2000 are  collected at 59 spatial locations scattered over 12 states in the Midwest of USA.  For ease of reference, we provide a map of Midwest states with the locations of the climate monitoring stations in Figure \ref{fig1}.
The Midwest  is a breadbasket of the United States and its agriculture has continued to play a major role in the economy of the region  \citep{Pryor2013}. The agriculture in the Midwest is vulnerably affected by the climate, of which precipitation is a vital component. To monitoring the future agricultural activities, it therefore has long been  recognized as an important problem to reveal how the change of precipitation takes place for different locations, different regions, or different years in the same region.
\vspace{-1cm}
\vspace{-5mm}

\begin{figure}[H]
	\begin{center}
		\begin{tabular}{cc}
			&\hspace{-7mm}	\includegraphics[height = 3.5cm, width = 4.3cm]{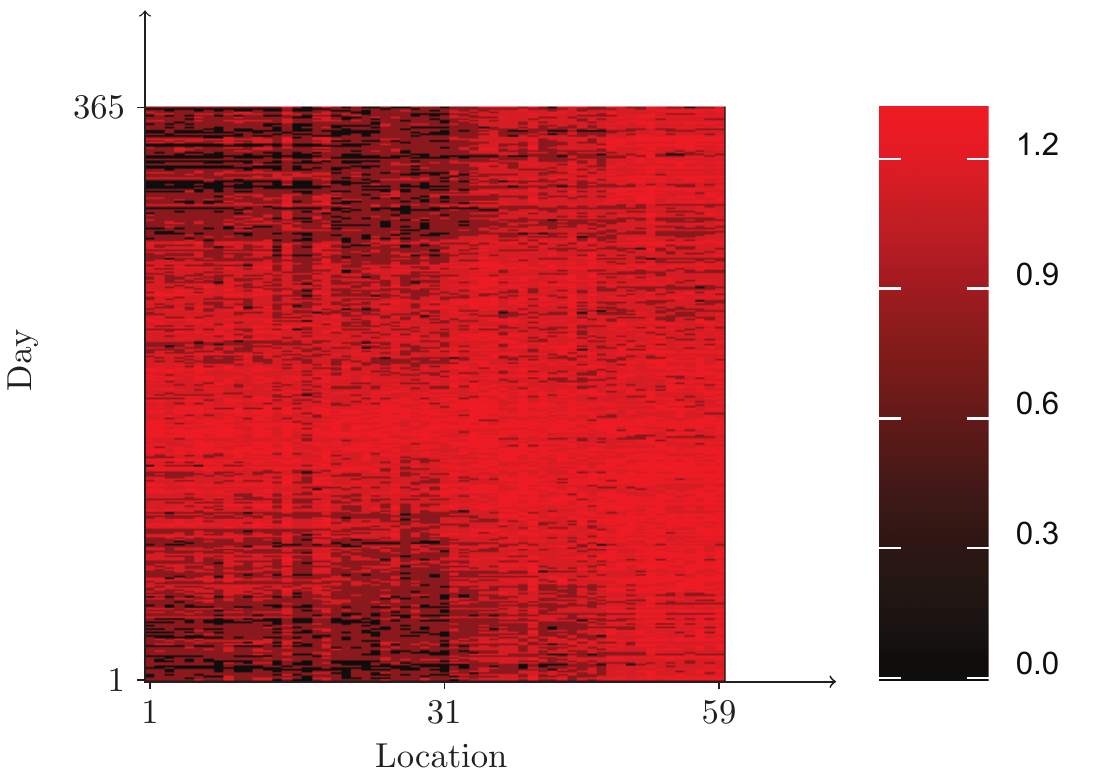}\hspace{2mm}\includegraphics[height = 4cm, width = 4cm]{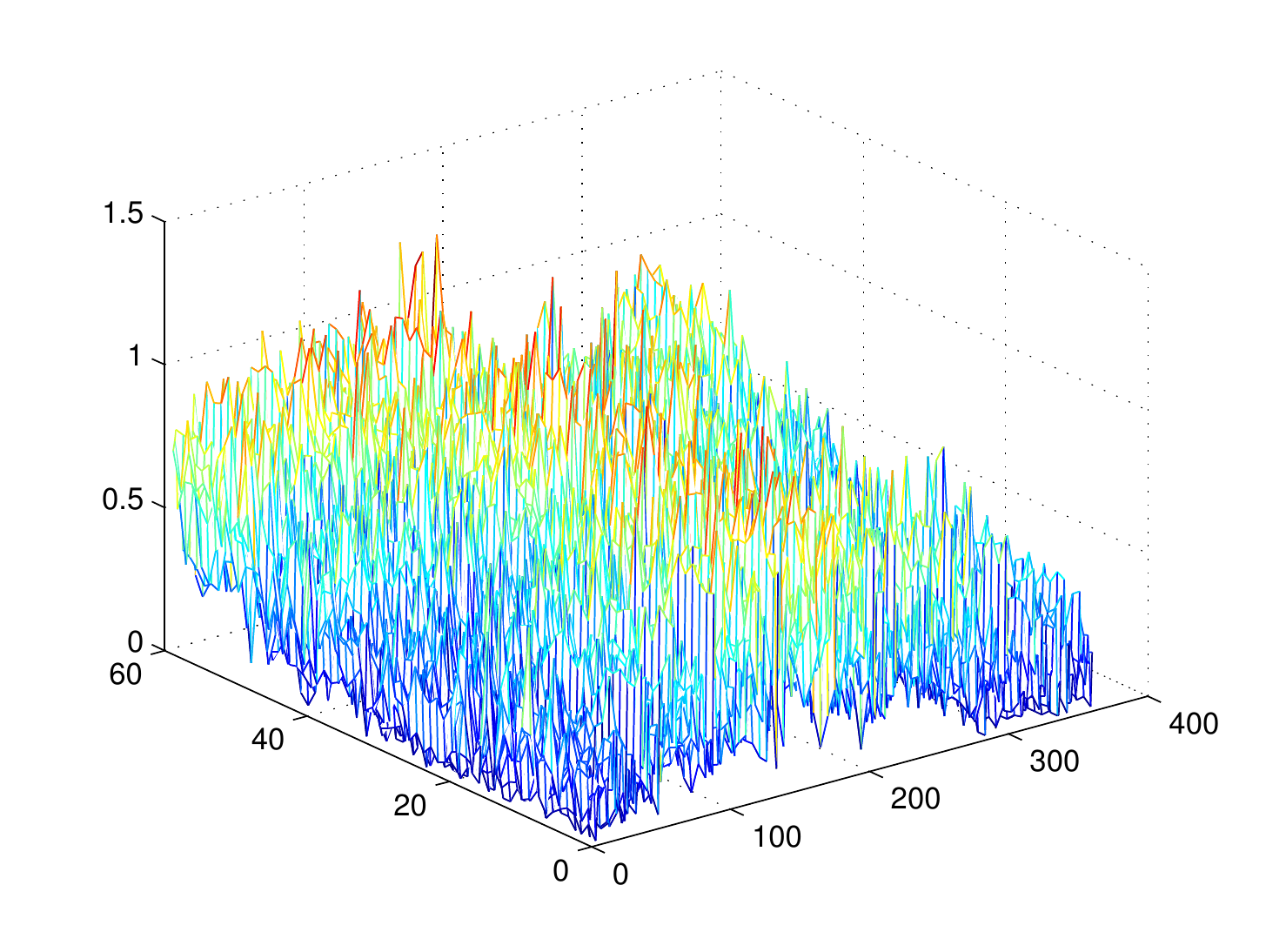}
		\end{tabular}
		\caption{The heatmap of sample mean surface of precipitation during the time 1941-1967 in the Midwest, where the first 31 stations are located in the Great Plains and the latter 28 stations are located in the Great Lakes. }\label{heatmap}
	\end{center}
\end{figure}

The study of the  precipitation  data has led to several interesting findings.  For instance, \cite{BerkesGabrys2009} detected no changes during the period 1941-2000 for only  individual station. However, it is difficult to implement if we sequently tested for every station when the number of stations were large.
\cite{GromenkoKokoszka2017JRSSb} used cumulative sum paradigm to expose the fact that, the mean precipitation curves before and after 1966 were different over the whole region. Nevertheless their method was particularly designed to detect the temporal change
but not applicable to detect the difference in spatio domain, not to mention
the joint spatiotemporal effect on the precipitation.
Looking into analysis of heatmaps of yearly sample mean surfaces where $X_{i}(s,t), i=1941,\cdots, 1967$,  corresponds to the  precipitation of the $t$th  day in the $i$th year
at the $s$th station,
intuitively we have observed that the yearly sample mean surface of precipitation in the Great Plains is different from that in the Great Lakes, refer to Figure \ref{heatmap}. Also, we can recognize from Figure \ref{heatmap} that some  profiles of mean surface are same but others are different.
These motivate us to develop more powerful inferential procedures to detect if mean surfaces or its profiles have significant difference for either different regions or different individual stations.

\vspace{-0.5cm}

Tracking back testing procedures for the equality of mean functions in the functional data setting, existing works mainly focus on detecting the curve equality for univariate functional data. 
In the two-sample testing scenario,
\cite{BenkoHardleKneip2009} presented bootstrap procedures for testing the equality of mean curves through the eigenelements for two independent functional samples.
Under the Gaussian assumption,
\cite{ZhangLiangXiao2010} considered the two-sample test based on $L^2$-norm.
\cite{FremdtHorvathKokoszkaSteinebach2014} derived mean functions comparison through a normal approximation method but only applicable to dense functional data samples.  
\cite{PomannStaicuGhosh2016} still solved testing the curve equality problem though in
bivariate (two-dimensional by their words) functional data setting and for distribution function testing.
Regarding the $k$-sample testing or the  one-way ANOVA for functional data, works include HANOVA \citep{FanLin1998}, Cram{\'e}r-von Mises type test \citep{CuevasFebreroFraiman2004, EstevezPerezVilar2013}, $F$-type test \citep{RamsaySilverman2005, Zhang2013, ZhangLiang2014},  B-spline test \citep{GoreckiSmaga2015}, and Mahalanobis distance \citep{GhigliettiLeva2017}, among others.
In the case of within-curve dependence in each sample,
\cite{AstonKirch2012MultiAnal} detected the mean curve variation using  $L^2$-norm criterion. \cite{StaicuLi2014} and its  multiple group extension  \cite{StaicuLahiriCarroll2015} worked on parametric testing relying on quite strong assumptions. Notice that, throughout our literature review, since our awareness concentrates on testing the equality of mean functions, we leave out other inferential topics such as testing the equality of coefficient operators or testing independency within a sample, and etc.

It has series of work in functional time series literature on testing the equality of mean functions, where weak dependence between or within two samples are accommodated in reality. Testing mean function difference in such functional time series study  had still been on comparison of mean curve functions (\citealp{ZhangShao2011, HorvathKokoszkaReeder2013, HorvathKokoszkaRice2014, HorvathRice2015RMC, HorvathRice2015JTSA, Torgovitski2015}, among others).

Aforementioned literature in both functional curve samples and functional time series have all inclined to testing the equality of mean curve functions, i.e. the inferential target is on univariate functional data.
However, for comparison between samples of multivariate functional data, there have been few works by far. Only \cite{GromenkoKokoszka2017JRSSb} raised testing the equality of the mean surfaces of bivariate functional data, but eventually the equality of mean curves indexed at all locations were tested. Also to the best of our knowledge, the profile test of mean surfaces has not been considered for two bivariate functional data samples.
Although the profile test of mean surfaces may belong to the curve test scope, it attributes to two different topics due to the different subjects. 
Above dire need in real-world data analysis and literature review motivates us to develop valid tests for equality of means surfaces and the corresponding profile test for  bivariate functional data samples.

To address the problem in demand, 
firstly, we obtain the marginal eigen-function of the pooled sample by marginal functional principal component analysis (FPCA) and project the profiles of mean surfaces on marginal eigenfunctions.  
The profile testing statistic measures the distance of the profile of mean surfaces for two bivariate functional samples. Once the marginal eigenfunctions are obtained,  
the eigensurfaces of the pooled sample can be constructed by further FPCA. 
The distance between mean surfaces for two samples can be measured by the globe test statistic using the analogous projection ideas.       
Consequently, our proposed profile testing procedures can be implemented for every profile of the mean surface, which corresponds to simultaneously test whether mean precipitation curves have significant difference for every station.  
The globe test performs well in terms of both the size and the power in that it includes the information of two domains effectively.

The major contribution of this paper is threefold.
Firstly, the presented methodology may be the first one to detect difference of mean surfaces and its profile for two-sample bivariate functional data.
In contrast to the literature that we can search out by far, of which the focus has almost all been on testing the equality of mean curves as a matter of fact. When one argument is fixed, our profile test methodology can also simultaneously detect the mean difference in the other domain. 
Secondly, our testing procedures are interpretable and easily implemented. This will help fill out some theoretical gaps in functional inference and facilitate the real application and interpretation in statistical perspective.
Finally, asymptotic distributions of the test statistics under null hypotheses has been derived. The consistency of  test procedure has been proved. 
In addition, simulation studies show that the proposed tests have a good control of the type I error by the size and can detect difference in mean surfaces and its profile effectively in terms of power in finite samples.


The rest of the paper is organized as follows.
In Section \ref{sec2}, we describe the model and data structure.
The profile test procedure of mean surfaces for two bivariate functional data samples is presented in Section \ref{sec3},
while globe test procedure is proposed in Section \ref{sec4}.
The finite sample performance for several representative scenarios is investigated in Section \ref{sec5}. In Section \ref{sec6}, we demonstrate two applications associated with the precipitation changes affected jointly by time and locations in the
Midwest of USA, and the trends in human mortality from European period life tables. The paper concludes with
a brief discussion in Section \ref{sec7}.   
Theory proofs are included in \hyperlink{App}{Supplementary}. 
\section{MODEL AND DATA STRUCTURE}\label{sec2}
Let $L^{2}(\mathcal{S} \times \mathcal{T})$ be the separable
Hilbert space.
$\{X^{(m)}(s,t):(s,t)\in \mathcal{S} \times \mathcal{T}\}$ is a square integrable  stochastic  process    on $L^{2}(\mathcal{S} \times \mathcal{T} )$  with
mean function $\mu_m(s,t)=E\{X^{(m)}(s,t)\}$ and covariance function
\begin{align*}
	C^{(m)}\{(s,v),(u,t)\}= \text{E}\{X^{(m)c}(s,v)X^{(m)c}(u,t)\},
\end{align*}
where $X^{(m)c}(s,t)=X^{(m)}(s,t)-\mu_m(s,t)$, for $m=1,2$, respectively.
With this notation, we
can decompose $X^{(m)}(s,t)$ into
$$\begin{array}{l}
X^{(m)}(s, t)=\mu_m(s, t)+\varepsilon^{(m)}(s,t),
m=1,2,
\end{array}$$
where $\varepsilon^{(m)}(s,t)$  is the stochastic part of $ X^{(m)}(s, t)$ with
$\text{E} \{\varepsilon^{(m)}(s,t)\}=0 $   and covariance function  $C^{(m)}\{(s,v),(u,t)\}$.

Functional samples $\{X_i^{(m)}(s, t),  m=1,2;  i=1,\cdots,n_m \}$ may usually be modeled as independent realizations of the underlying  stochastic process $X^{(m)}(s,t)$.
In practice,  $\{X_i^{(m)}(s, t),  m=1,2;  i=1,\cdots,n_m \}$  can not be observed, but rather,  measurements are taken at discrete  time points.   In this paper,   we assume  $\{X_i^{(m)}(s, t),  m=1,2;  i=1,\cdots,n_m \}$
are recorded on a regular and dense grid of  time points as follows,
$$\begin{array}{l}
X_i^{(m)}(s_{il_1}, t_{il_2})=\mu_m(s_{il_1}, t_{il_2})+\varepsilon_i^{(m)}(s_{il_1},t_{il_2});\vspace{1mm}\\
m=1,2;\ i=1,\cdots,n_m;\ l_1=1,\cdots,N;\ l_2=1,\cdots,M.
\end{array}$$

In this paper, we are  firstly interested in profile test of bivariate functional data samples,  i.e.  for every fixed $t^{*} \in \mathcal{T}$,
\begin{eqnarray}\label{2.1a}
	\begin{aligned}
		&H_{0}^{\mathcal{S}}: \mu_{1}(s,t^{*})=\mu_{2}(s,t^{*}) \hspace{4pt}  \text{vs.} \hspace{4pt}
		H_{1}^{\mathcal{S}}: \mu_{1}(s,t^{*})\neq\mu_{2}(s,t^{*}),\hspace{4pt} s \in \mathcal{S},
	\end{aligned}
\end{eqnarray}
or for every fixed $s^{*} \in \mathcal{S}$,
\begin{eqnarray}\label{2.1b}
	\begin{aligned}
		&H_{0}^{\mathcal{T}}: \mu_{1}(s^{*},t)=\mu_{2}(s^{*},t) \hspace{4pt}  \text{vs.} \hspace{4pt}
		H_{1}^{\mathcal{T}}: \mu_{1}(s^{*},t)\neq\mu_{2}(s^{*},t),\hspace{4pt} t \in \mathcal{T}.
	\end{aligned}
\end{eqnarray}

Then we go to the second target to present a globe test procedure for bivariate functional data samples  with hypothesis  below,
\begin{eqnarray}\label{2.2}
	\begin{aligned}
		&H_{0}: \mu_{1}(s,t)=\mu_{2}(s,t) \hspace{4pt}  \text{vs.} \hspace{4pt}
		H_{1}: \mu_{1}(s,t)\neq\mu_{2}(s,t),\hspace{4pt} s \in \mathcal{S},  t \in \mathcal{T}.
	\end{aligned}
\end{eqnarray}

The equality in hypothesis \eqref{2.1a} means that $\int_{\mathcal{S}} \{\mu_{1}(s,t^{*})-\mu_{2}(s,t^{*})\}^2 ds =0$   for every fixed $t^{*} \in \mathcal{T}$, and the alternative means that $\int_{\mathcal{S}}\{\mu_{1}(s,t^{*})-\mu_{2}(s,t^{*})\}^2 ds >0$.   Analogously meaning can be interpreted for \eqref{2.1b}. However,  null hypothesis of \eqref{2.2} implies $\int_{\mathcal{S}}\int_{\mathcal{T}}\{\mu_{1}(s,t)-\mu_{2}(s,t)\}^2dtds=0$ while the alternative means that $\int_{\mathcal{S}}\int_{\mathcal{T}}\{\mu_{1}(s,t)-\mu_{2}(s,t)\}^2dtds>0$.
For statistical inference of bivariate functional data,  marginal FPCA is a  widely used tool, which often assumes that  bivariate functional data can project onto  finite-dimensional eigensurfaces \citep{LiGuan2014,ParkStaicu2015, AstonPigoliTavakoli2017}. It is our start point for the proposed profile and globe test procedures.

\section{Profile test of bivariate functional data}\label{sec3}
Profile test of bivariate functional data is an important problem, as it allows to
provide multiple insight from multiple angles, and also is of interest in many
applications. For example, in analysis of precipitation, the testing problem \eqref{2.1b} corresponds to test whether mean precipitation curves have significant difference  before and after 1966 for every station, while the testing problem \eqref{2.1a} means to test  whether different stations have significant difference for every day. \cite{BerkesGabrys2009} considered detection the difference only on an individual station. However, it is difficult to implement when the number of stations is large if we sequentially test for every station by their method. 
So, we propose the profile test of mean functions which is easy to implement and can  simultaneously detect difference of all stations. In this section, we address the test problem \eqref{2.1a} only as \eqref{2.1b} can be analogously implemented.

As a first step, the marginal covariance function is denoted to be $G_\mathcal{S}^{(m)}(s,u)=\int_\mathcal{T}C^{(m)}\{(s,t),(u,t)\}dt$, as the form of (5) in \cite{ChenDelicadoMuller2017}, and may be estimated by
\begin{eqnarray}
	\begin{aligned}
		\widehat{G}_{\mathcal{S}}^{(m)}(s_h,s_l)={\displaystyle\frac{1}{n_mM}}\sum\limits_{i=1}^{n_m}\sum\limits_{k_2=1}^{M}
		&\widehat{X}_{i}^{(m)c}(s_h,t_{ik_2})\\
		&\times\widehat{X}_{i}^{(m)c}(s_l,t_{ik_2}),
	\end{aligned}
\end{eqnarray}
where $\widehat{X}_{i}^{(m)c}(s, t)=X_{i}^{(m)}(s,t)-\overline{ X}^{(m)}(s,t)$
with \\ $\overline{ X}^{(m)}(s,t)={\displaystyle\frac{1}{n_m}}\sum\limits_{i=1}^{n_m}
{X}_{i}^{(m)}(s,t)$.

Denote
\[
\begin{array}{ll}
\widehat{G}_{\mathcal{S}}(s,u)={\displaystyle\frac{n_2}{n_1+n_2}}\widehat{G}_{\mathcal{S}}^{(1)}(s,u)
+{\displaystyle\frac{n_1}{n_1+n_2}}\widehat{G}_{\mathcal{S}}^{(2)}(s,u),\ s,u\in \mathcal{S}.
\end{array}
\]

It is easy to see  $\widehat{G}_{\mathcal{S}}(s,u) \overset{p}{\longrightarrow} (1-\theta)G_{\mathcal{S}}^{(1)}(s,u)+\theta G_{\mathcal{S}}^{(2)}(s,u)\equiv G_{\mathcal{S}}(s,u)$, where  $\theta$ is defined in Assumption 6 stated in next section and  $G_{\mathcal{S}}(s,u)$ is the pooled covariance function.
Consequently, it has orthogonal eigenfunctions $\{\psi_{j}\}_{j\geq 1}$ and
non-negative eigenvalues $\{\nu_{j}\}_{j\geq 1}$ satisfying
\[
\begin{array}{ll}
&\displaystyle \int_{\mathcal{S}}^{} G_{\mathcal{S}}(s,u)\psi_{j}(u)du=\nu_{j}\psi_{j}(s),\  s,u\in \mathcal{S},\ j=1,2,\ldots.
\end{array}
\]

Such eigencomponents can be numerically estimated by suitably discretized eigenequations,
\begin{eqnarray}\label{3.2}
	\displaystyle \int_{\mathcal{S}}\widehat{G}_{\mathcal{S}}(s,u)\widehat{\psi}_{j}(u)du=
	\widehat{\nu}_{j}\widehat{\psi}_{j}(s),\ j=1,2,\ldots,
\end{eqnarray}
with orthogonal constraints on $\{\widehat{\psi}_{j}\}_{j\geq 1}$.

Once the estimators of marginal eigen-functions $\widehat{\psi}_{j}(s)$, $j=1,2,\ldots,$  are obtained, we project the observations onto the marginal eigenfunctions and obtain the
profile estimators of mean functions as follows:  for every fixed $t^{*} \in \mathcal{T}$,

\begin{eqnarray}\label{3.3}
	&\widehat{\mu}_{m}(s,t^{*})=\sum\limits_{j=1}^{J}\widehat{\eta}_{j}^{(m)}(t^{*})
	\widehat{\psi}_{j}(s),\ m=1,2, 
\end{eqnarray}
with
\begin{eqnarray*}
	\begin{aligned}
		&\widehat{\eta}_{j}^{(m)}(t^{*})={\displaystyle\frac{1}{n_m}}\sum\limits_{i=1}^{n_m}
		\widehat{\eta}_{ij}^{(m)}(t^{*}),\\
		&\widehat{\eta}_{ij}^{(m)}(t^{*})={\displaystyle\frac{1}{N}}
		\sum\limits_{l_1=1}^{N}X_{i}^{(m)}(s_{il_{1}},t^{*})
		\widehat{\psi}_{j}(s_{il_{1}}).
	\end{aligned}
\end{eqnarray*}
For practical implementation, one has to decide the magnitude of $J$.
A practical strategy is $J=\min\{j:\frac{\widehat{\nu}_{1}+\widehat{\nu}_{2}+\cdots
	+\widehat{\nu}_{k}}{\widehat{\nu}_{l}+\widehat{\nu}_{2}+\cdots}>q\}$,  where  $\widehat{\nu}_{l}$, $l=1,2,\cdots$  are defined in \eqref{3.2}.
We find that  $q=90\%$ threshold works well for our numerical examples.

Based on  above discussion, we propose the following profile test statistic
\begin{eqnarray*}
	\widehat{\text{TP}}(t^{*})={\displaystyle\frac{n_1n_2}{n_1+n_2}}\sum\limits_{j=1}^{J}
	{\displaystyle\frac{\left(\widehat{\eta}_{j}^{(1)}(t^{*})- \widehat{\eta}_{j}^{(2)}(t^{*})\right)^{2}}{\widehat{\lambda}_{j}(t^{*})}},
\end{eqnarray*}
where
$\widehat{\lambda}_{j}(t^{*})=\frac{n_2}{n_1+n_2}\widehat{\lambda}_{j}^{(1)}(t^{*})
+\frac{n_1}{n_1+n_2}\widehat{\lambda}_{j}^{(2)}(t^{*})$ with\\
$\widehat{\lambda}_{j}^{(m)}(t^{*})=n_m^{-1}\sum_{i=1}^{n_m}
\left\{\widehat{\eta}_{ij}^{(m)}(t^{*})-\widehat{\eta}_{j}^{(m)}(t^{*})\right\}^{2}$, $m=1,2$.

\begin{remark}\label{rem1}
	It is easy to see that $\frac{n_1n_2}{n_1+n_2}\int[\widehat{\mu}_{1}(s,t^{*})$ \\ $- \widehat{\mu}_{2}(s,t^{*})]^{2}dt \overset{p}{\longrightarrow} U_{n_1, n_2}= \frac{n_1n_2}{n_1+n_2} \sum_{l=1}^{ K}(\widehat{\eta}_{l}(t^{*})- \widehat{\eta}_{2}(t^{*}))^{2}$. However,  the variance of $U_{n_1, n_2}$ may be unnecessarily inflated by the
	presence of, possibly many, very small estimates $\widehat{\mu}_{1}(s,t^{*})- \widehat{\mu}_{2}(s,t^{*})$.  This drawback can be remedied by giving a divisor to $\widehat{\lambda}_{j}(t^{*})$.
\end{remark}

We then establish asymptotic behaviors of the test statistic $\widehat{\text{TP}}(t^{*})$ under the null hypothesis $H_{0}^{\mathcal{S}}$ and the alternative one $H_{1}^{\mathcal{S}}$. To derive the asymptotic properties of profile test statistic, we make the following assumptions.

\begin{assumption}\label{ass1}
	$\nu_{1}>\nu_{2}>\cdots$
	where $\{\nu_{j}\}_{j=1,2,\ldots}$ are the eigenvalues of covariance operates  $G_{S}(s,u)$.
\end{assumption}

\begin{assumption}\label{ass2}
	For every fixed $t^*$,   $\mu_{m}(s,t^*), m=1,2$ may be written as $\mu_{m}(s,t^*)=\sum_{j=1}^\infty
	\eta_{j}^{(m)}(t^*)\psi_{j}(s)$,  where $\eta_{j}^{(m)}(t^*)=\int_{0}^{1}\mu_{m}(s,t^*)\psi_{j}(s)ds$.
\end{assumption}

\begin{assumption}\label{ass3}
	Assume $\sup _{(s,t)\in \mathcal{S} \times \mathcal{T}}\mu_{m}^{2}(s,t), m=1,2$   are bounded
	and $\text{E}(\sup|\varepsilon^m(s,t)|^{4}), m=1,2$ are bounded.
\end{assumption}

\begin{assumption}\label{ass4}
	The grid point $\{t_{il_1}:l_1=1,\ldots,N\}$  and
	$\{s_{il_2}: l_2=1,\ldots,M\}$ are equidistant. We assume
	$n_1/N^{2}=o(1)$,  $n_1/M^{2}=o(1)$,  $n_2/N^{2}=o(1)$   and $n_2/M^{2}=o(1)$.
\end{assumption}

\begin{assumption}\label{ass5}
	$\min\{n_1,n_2\}\rightarrow\infty$, $n_1/(n_1+n_2)\rightarrow\theta$  for a fixed constant $\theta\in (0,1)$.
\end{assumption}

Assumptions \ref{ass1} and \ref{ass3} are regular conditions. One needs these conditions to uniquely (up to signs) choose $\psi_{j}(s)$ and obtain  the bound of $\widehat{\psi}_{j}(s)-\psi_{j}(s)$.  Assumption \ref{ass2} means that the profiles of mean surface are projected onto a space that is generated by a large set of basis functions.  
Assumption \ref{ass4} requires that functional data are recorded on dense grid.
Assumption \ref{ass5} is of standard for two-sample asymptotic inference.

\begin{theorem}\label{thm1}
	Under Assumptions \ref{ass1}-\ref{ass5} and $H_0^{\mathcal{S}}$, we have $\widehat{\mathrm{TP}}(t^{*})\overset{d}{\longrightarrow}\chi_{J}^{2}$,
	where $\chi_{J}^{2}$ stands for a $\chi^{2}$ -distributed random variable with $J$ degrees of freedom. Under $H_1^{\mathcal{S}}$ and $0<\theta<1$, we have $\widehat{\mathrm{TP}}(t^{*})\overset{p}{\longrightarrow}\infty$.
\end{theorem}

From the expression of $\widehat{\mathrm{TP}}(t^{*})$ and remark \ref{rem1}, we can see that $\widehat{\mathrm{TP}}(t^{*})$ depends on sample sizes $n_1, n_2$, and $\widehat{\eta}_{j}^{(1)}(t^{*})- \widehat{\eta}_{j}^{(2)}(t^{*}), j=1,\cdots,J$,  which
reflects  the difference of profile mean functions $\mu_1(s,t^*)$ and $\mu_2(s,t^*)$.
Intuitively, $\sqrt{\frac{n_1n_2}{n_1+n_2}}{(\widehat{\eta}_{j}^{(1)}(t^{*})- \widehat{\eta}_{j}^{(2)}(t^{*}))}\widehat{\lambda}_{j}^{-1/2}$  has a limiting standard  normal distribution under  $H_{0}^{\mathcal{S}}$.
Theorem \ref{thm1} shows that $\widehat{\text{TP}}(t^{*})$ asymptotically follows the chi-square distribution with $J $ degrees of freedom if $H_0^{\mathcal{S}}$ holds. Furthermore, $\widehat{\text{TP}}(t^{*})$ is consistent under $H_1^{\mathcal{S}}$. The proof of this theorem is provided in \hyperlink{App}{Supplementary}.

\section{Globe test of bivariate functional data}\label{sec4}
Compared with the profile test, the globe test of bivariate functional data attempts to detect the joint effects impacted by both domains. In this section, we develop a globe test method for bivariate functional data which aims to detect whether mean surfaces of precipitation have significant difference over a specific time window and/or a specific area, or whether two regions exist significant difference during different time windows.

Based on the estimated  marginal eigenfunctions $\widehat{\psi}_{j}(s)$  in Section \ref{sec3}, we next estimate the
marginal  functional  principal component scores
$\widetilde{\xi}^{(m)}_{j,i}(t)$.
The traditional integral estimates of
$\widetilde{\xi}^{(m)}_{j,i}(t)$  based on the definition
\begin{eqnarray*}
	\begin{aligned}
		&\widetilde{\xi}^{(m)}_{j,i}(t)=\int_{\mathcal{S}} X_{i}^{(m)c}(s,t)\widehat{\psi}_{j}(s)ds,\\
		&i=1,\ldots,n_m;\ j=1,2,\ldots. 
	\end{aligned}
\end{eqnarray*}
are
\begin{eqnarray}\label{4.1}
	\begin{aligned}
		&\widehat{\xi}^{(m)}_{j,i}(t)=\sum_{l=2}^{N} X_{i}^{(m)c}(s_{l},t)\widehat{\psi}_{j}(s_{l})(s_{l}-s_{l-1}),\\ &i=1,\ldots,n_m;\ j=1,2,\ldots.
	\end{aligned}
\end{eqnarray}
where  $N$  is the number of measurements for  $X_{i}^{(m)c}(s,t)$  in the direction $\mathcal{S}$.

Notice that each score function $\widehat{\xi}^{(m)}_{j,i}(t)$ is a centered new random curve. Denote the covariance function of $\xi_{j,i}^{(m)}(t)$ by $G_{\mathcal{T},j}^{(m)}(v,t)=E\{\xi_{j,i}^{(m)}(v)\xi_{j,i}^{(m)}(t)\}$. Then,  the  estimator of $G_{\mathcal{T},j}^{(m)}$ is denoted as,
\begin{eqnarray*}
	\begin{aligned}
		&\widehat{G}^{(m)}_{\mathcal{T},j}(t_h,t_l)={\displaystyle\frac{1}{n_m}}\sum\limits_{i=1}^{n_m}
		\widehat{\xi}_{j,i}^{(m)}(t_h)\widehat{\xi}_{j,i}^{(m)}(t_l),\\ 
		&t_h, t_l\in \mathcal{T}; \ j=1,2,\ldots. 
	\end{aligned}
\end{eqnarray*}
Let
\begin{eqnarray*}
	\begin{aligned}
		&\widehat{G}_{\mathcal{T},j}(v,t)={\frac{n_2}{n_1+n_2}}\widehat{G}^{(1)}_{\mathcal{T},j}(v,t)
		+{\displaystyle\frac{n_1}{n_1+n_2}}\widehat{G}^{(2)}_{\mathcal{T},j}(v,t),\\
		&v,t\in \mathcal{T};\ j=1,2,\ldots.
	\end{aligned}
\end{eqnarray*}
It is easy to see $\widehat{G}_{\mathcal{T},j}(v,t)\overset{p}{\longrightarrow}
(1-\theta)G^{(1)}_{\mathcal{T},j}(v,t)+\theta G^{(2)}_{\mathcal{T},j}(v,t)$\\ $\equiv G_{\mathcal{T},j}(v,t)$
where
$G_{\mathcal{T},j}(v,t)$ is the covariance function   and  has orthogonal eigenfunctions $\{\phi_{jk}\}_{k\geq 1}$ and
non-negative eigenvalues $\{\nu_{jk}\}_{k\geq 1}$ satisfying
\begin{eqnarray*}
	\begin{aligned}
		&\int_{\mathcal{T}}  G_{\mathcal{T},j}(v,t)\phi_{jk}(v)dv=\nu_{jk}\phi_{jk}(t),\\
		&v,t\in \mathcal{T};\ k, j=1,2,\ldots.
	\end{aligned}
\end{eqnarray*}

Then estimators of eigenvalues and eigenfunctions $\{(\nu_{jk}, \phi_{jk}(t)): j,k\geq 1\}$ are obtained by the following equations,
\begin{eqnarray}\label{4.2}
	\displaystyle\int_{\mathcal{T}}\widehat{G}_{\mathcal{T},j}(v,t)\widehat{\phi}_{jk}(v)dv=
	\widehat{\nu}_{jk}\widehat{\phi}_{jk}(t),\ k, j=1,2,\ldots, 
\end{eqnarray}
with orthogonal constraints on $\{\widehat{\phi}_{jk}\}_{k\geq 1}$.

Denote $\varphi_{jk}(s,t)\equiv\phi_{jk}(t)\psi_{j}(s)$ and its consistent estimator by $\widehat{\varphi}_{jk}(s,t)=\widehat{\phi}_{jk}(t)\widehat{\psi}_{j}(s)$. We propose estimators of the mean surfaces which
are projection of observations onto a hyperspace spanned from the pooled eigensurfaces $\{\widehat{\varphi}_{jk}(s,t): j,k\geq 1\}$, written as
\begin{eqnarray}\label{4.3}
	\widehat{\mu}_{m}(s,t)=\sum\limits_{j=1}^{J}\sum\limits_{k=1}^{K_{j}}\widehat{\eta}_{jk}^{(m)}
	\widehat{\varphi}_{jk}(s,t),\ m=1,2, 
\end{eqnarray}
with
\begin{eqnarray*}
	\begin{aligned}
		&\widehat{\eta}_{jk}^{(m)}={\displaystyle\frac{1}{n_m}}\sum\limits_{i=1}^{n_m}\widehat{\eta}_{ijk}^{(m)},\\
		&\widehat{\eta}_{ijk}^{(m)}={\displaystyle\frac{1}{MN}}
		\sum\limits_{l_2=1}^{M}\sum\limits_{l_1=1}^{N}X_{i}^{(m)}(s_{il_{1}},t_{il_{2}})\widehat{\varphi}_{jk}(s_{il_{1}},t_{il_{2}}),
	\end{aligned}
\end{eqnarray*}
where selection of $J$ is the same to in Section \ref{sec3} and $K_j$ can be decided by analogous procedure. In details, we select  $K_j=\min\{k:\frac{\widehat{\nu}_{j1}+\widehat{\nu}_{j2}+\cdots
	+\widehat{\nu}_{jk}}{\widehat{\nu}_{j1}+\widehat{\nu}_{j2}+\cdots}>0.9\}$, where $\widehat{\nu}_{jl}$, $l=1,2,\cdots$ are defined in \eqref{4.2}.

It is natural to take into consideration the term $\widetilde{\text{TC}}\equiv \int_{\mathcal{S}}\int_{\mathcal{T}}\{\mu_{1}(s,t)-\mu_{2}(s,t)\}^2dtds$ to measure the distance between two estimated mean surfaces.
It is readily seen that  $\widetilde{\text{TC}}\overset{p}{\longrightarrow}\sum\limits_{j=1}^{J}\sum\limits_{k=1}^{K_{j}}\left(\widehat{\eta}_{jk}^{(1)}- \widehat{\eta}_{jk}^{(2)}\right)^2 $.
Therefore, $H_{0}$  will be rejected if  $\widetilde{\text{TC}}$ is large.  Similarly,  the variance of $\widetilde{\text{TC}}$ may be unnecessarily inflated by the
presence of, possibly many, very small estimates $\widehat{\eta}_{jk}^{(1)}- \widehat{\eta}_{jk}^{(2)}$. This drawback can also be remedied by giving a divisor to their variance.

Based on the above steps, we propose the following test statistic
\[
\begin{array}{ll}
\widehat{\text{TM}}={\displaystyle\frac{n_1n_2}{n_1+n_2}}\sum\limits_{j=1}^{J}
\sum\limits_{k=1}^{K_{j}}{\displaystyle\frac{\left(\widehat{\eta}_{jk}^{(1)}- \widehat{\eta}_{jk}^{(2)}\right)^{2}}{\widehat{\lambda}_{jk}}},
\end{array}
\]
where
$\widehat{\lambda}_{jk}=n_2(n_1+n_2)^{-1}\widehat{\lambda}_{jk}^{(1)}
+n_1(n_1+n_2)^{-1}\widehat{\lambda}_{jk}^{(2)}$  with
$\widehat{\lambda}_{jk}^{(m)}=(n_m-1)^{-1}\sum_{i=1}^{n_m}
\left(\widehat{\eta}_{ijk}^{(m)}-\widehat{\eta}_{jk}^{(m)}\right)^{2}$, $m=1,2$.

From \eqref{4.3}, we can see that $X_{i}^{(1)}(\cdot,\cdot)$ and  $X_{i}^{(2)}(\cdot,\cdot)$
are directly projected on  the common basis surface and obtain $\widehat{\eta}_{ijk}^{(1)}$ and $\widehat{\eta}_{ijk}^{(2)}$.  $\widehat{\eta}_{jk}^{(1)}$ and $\widehat{\eta}_{jk}^{(2)}$, which are the average of such projection, and hence can be viewed as the scores of projection that two  mean surfaces $\mu_{1}(s,t)$   and $\mu_{2}(s,t)$  project on the same basis function space, respectively.
The representation of $\widehat{\text{TM}}$ measures the total such deviation between two samples. Therefore, the proposed method has a nice explanation and easy to implement.

Next we establish asymptotic behavior of the test statistic $\widehat{\text{TM}}$ under hypotheses (\ref{2.2}). Additionally, we need the following assumptions.

\begin{assumption}\label{ass6}
	$\nu_{j1}>\nu_{j2}>\cdots$ where $\{\nu_{jk}\}_{k=1,2,\ldots; j=1,2,\ldots}$ are the eigenvalues of the covariance function $G_{T}(v,t)$.
\end{assumption}

\begin{assumption}\label{ass7}
	Assume $\mu_{m}(s,t), m=1,2$  may be written as $\mu_{m}(s,t)=\sum_{j=1}^\infty
	\sum_{k=1}^{\infty}\eta_{jk}^{(m)}\varphi_{jk}(s,t)$,  where $\eta_{jk}^{(m)}=\int_{0}^{1}\int_{0}^{1}\mu_{m}(s,t)\varphi_{jk}(s,t)dsdt$.
\end{assumption}

Assumption \ref{ass6} along with Assumption \ref{ass3} in Section \ref{sec3} ensures the bound of $\widehat{\phi}_{jk}(t)-\widehat{\phi}_{jk}(t)$. The interpretation of Assumption \ref{ass7} is similar to Assumption \ref{ass2} in Section \ref{sec3}. 

\begin{theorem}\label{thm2}
	Under Assumptions \ref{ass1}-\ref{ass7} and $H_{0}$,  we have
	\[
	\begin{array}{llll}
	\widehat{\mathrm{TM}}\overset{d}{\longrightarrow}\chi_{\sum\limits_{j=1}^{J}K_{j}}^{2},
	\end{array}
	\]
	where $\chi_{\sum_{j=1}^{J}K_{j}}^{2}$ stands for a $\chi^{2}$-distributed random variable with $\sum_{j=1}^{J}K_{j}$ degrees of freedom. Under $H_1$ and $0<\theta<1$, we have $\widehat{\mathrm{TM}}\overset{p}{\longrightarrow}\infty$.
\end{theorem}

Intuitively $\sqrt{\frac{n_1n_2}{n_1+n_2} }\left(\widehat{\eta}_{jk}^{(1)}- \widehat{\eta}_{jk}^{(2)}\right)\widehat{\lambda}_{jk}^{-1/2}$  has a limiting standard  normal distribution under $H_{0}$. Theorem \ref{thm2} shows that $\widehat{\text{TM}}$ asymptotically follows the chi-square distribution with $\sum\limits_{j=1}^{J}K_{j} $ degrees of freedom under $H_0$.
The consistency of  $\widehat{\text{TM}}$  is also illustrated under $H_1$, which
together provides clear theoretical justification of the empirical properties of the proposed test.
The proof of this theorem is provided in \hyperlink{App}{Supplementary}.
\section{Simulation studies}\label{sec5}
We conduct extensive simulation studies and report two representative examples here. Examples \ref{eg1} and \ref{eg2} evaluate two proposed testing procedures in terms of empirical size and power when covariance functions of two samples are identical or distinct, separately. The data grid for argument $s$ consists of 100 equispaced points on $[0, 1]$, and the grids for argument $s$ consists of 50 equispaced points on $[0,1]$. Each pair of data-generated processes was replicated 1000 times.
\begin{example}\label{eg1}
	Identical covariance functions.
\end{example}
\noindent In this example, we consider the following model
\begin{eqnarray}\label{5.1}
	\begin{aligned}
		&X_{i}^{(1)}(s, t)=\varepsilon_{i}^{(1)}(s, t),\ i=1,\ldots,n_1,\\
		&X_{i}^{(2)}(s, t)=\delta (s+t)+ \varepsilon_{i}^{(2)}(s, t),\ i=1,\ldots,n_2,
	\end{aligned}
\end{eqnarray}
where $\varepsilon_{i}^{(1)}(s, t)$  and  $\varepsilon_{i}^{(2)}(s, t)$ are independently generated from
\[\varepsilon(s, t)=\sum\limits_{j=1}^{2}\xi_{j}(t)\psi_{j}(s),\ s\in [0,1],t\in [0,1],\]
with $\psi_{1}(s)=s^{2}$ and $\psi_{2}(s)=s^{3}$, $s\in [0,1]$.
$\xi_{j}(t)$ is generated from
\[
\begin{array}{llll}
\xi_{j}(t)=\sum\limits_{k=1}^2\chi_{jk}\phi_{jk}(t),\ j=1,2,
\end{array}
\]
with
$\phi_{11}(t)=\phi_{21}(t)=-\surd2\cos(2\pi t)$, $\phi_{12}(t)=\phi_{22}(t)=\surd2\sin(2\pi t)$, $t\in [0,1]$;
$\chi_{11}\sim N(0,3)$, $\chi_{12}\sim N(0,1.5)$, $\chi_{21}\sim N(0,2)$,  and  $\chi_{22}\sim N(0,1)$.

\begin{example}\label{eg2}
	Distinct covariance functions.
\end{example}
\noindent To compare with Example \ref{eg1}, we consider the following model
\begin{eqnarray}\label{5.2}
	\begin{aligned}
		&X_{i}^{(1)}(s, t)=\varepsilon_{i}^{(1)}(s, t),\ i=1,\ldots,n_1,\\
		&X_{i}^{(2)}(s, t)=\delta (s+t)+ \varepsilon_{i}^{(2)}(s, t),\ i=1,\ldots,n_2,
	\end{aligned}
\end{eqnarray}
where $\varepsilon_{i}^{(1)}(s, t)$ is generated from
\[
\begin{array}{llll}
\varepsilon^{(1)}(s, t)=\sum_{j=1}^{2}\xi_{j}(t)\psi_{j}(s),\ s\in [0,1],t\in [0,1],
\end{array}
\]
and  $\varepsilon_{i}^{(2)}(s, t)$ from
\[
\begin{array}{llll}
\varepsilon^{(2)}(s, t)=\xi_{1}(t)\psi_{1}(s),\ s\in [0,1],t\in [0,1],
\end{array}
\]
with $\psi_{1}(s)=s^{2}$ and $\psi_{2}(s)=s^{3}$, $s\in [0,1]$.
$\xi_{j}(t)$ is generated from
\[
\begin{array}{llll}
\xi_{j}(t)=\sum\limits_{k=1}^2\chi_{jk}\phi_{jk}(t),\ j=1,2,
\end{array}
\]
with
$\phi_{11}(t)=\surd2\cos(2\pi t)$,
$\phi_{21}(t)=\surd2\sin(2\pi t)$,
$\phi_{12}(t)=2\cos(4\pi t)$, $\phi_{22}(t)=2\sin(4\pi t)$, $t\in [0,1]$;
$\chi_{11}\sim N(0,3)$, $\chi_{12}\sim N(0,1.5)$, $\chi_{21}\sim N(0,2)$,  and  $\chi_{22}\sim N(0,1)$.

\begin{figure}[H]
	\begin{center}
		\begin{tabular}{cc}
			\includegraphics[height = 3.5cm, width = 4cm]{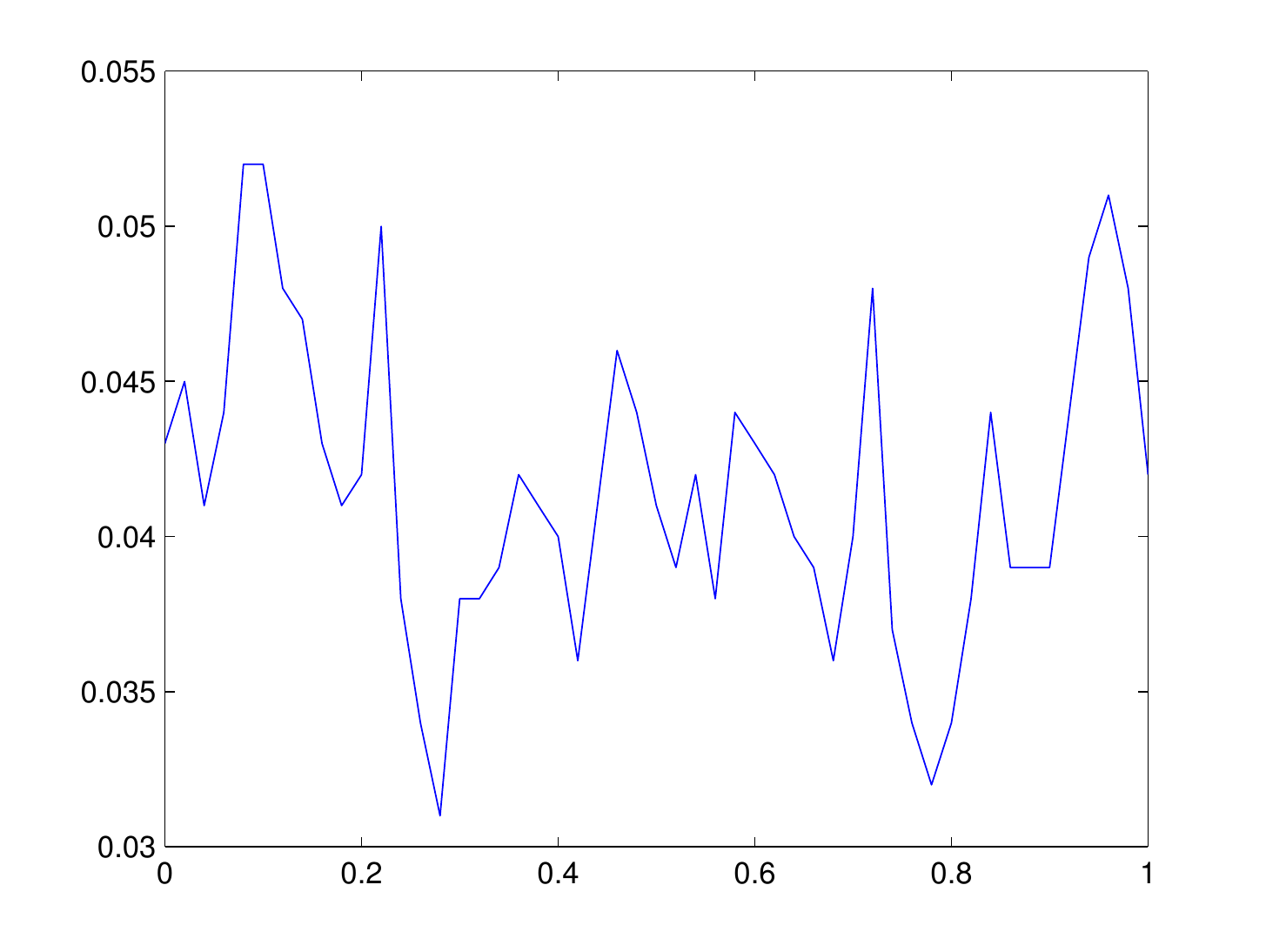}
			\includegraphics[height = 3.5cm, width = 4cm]{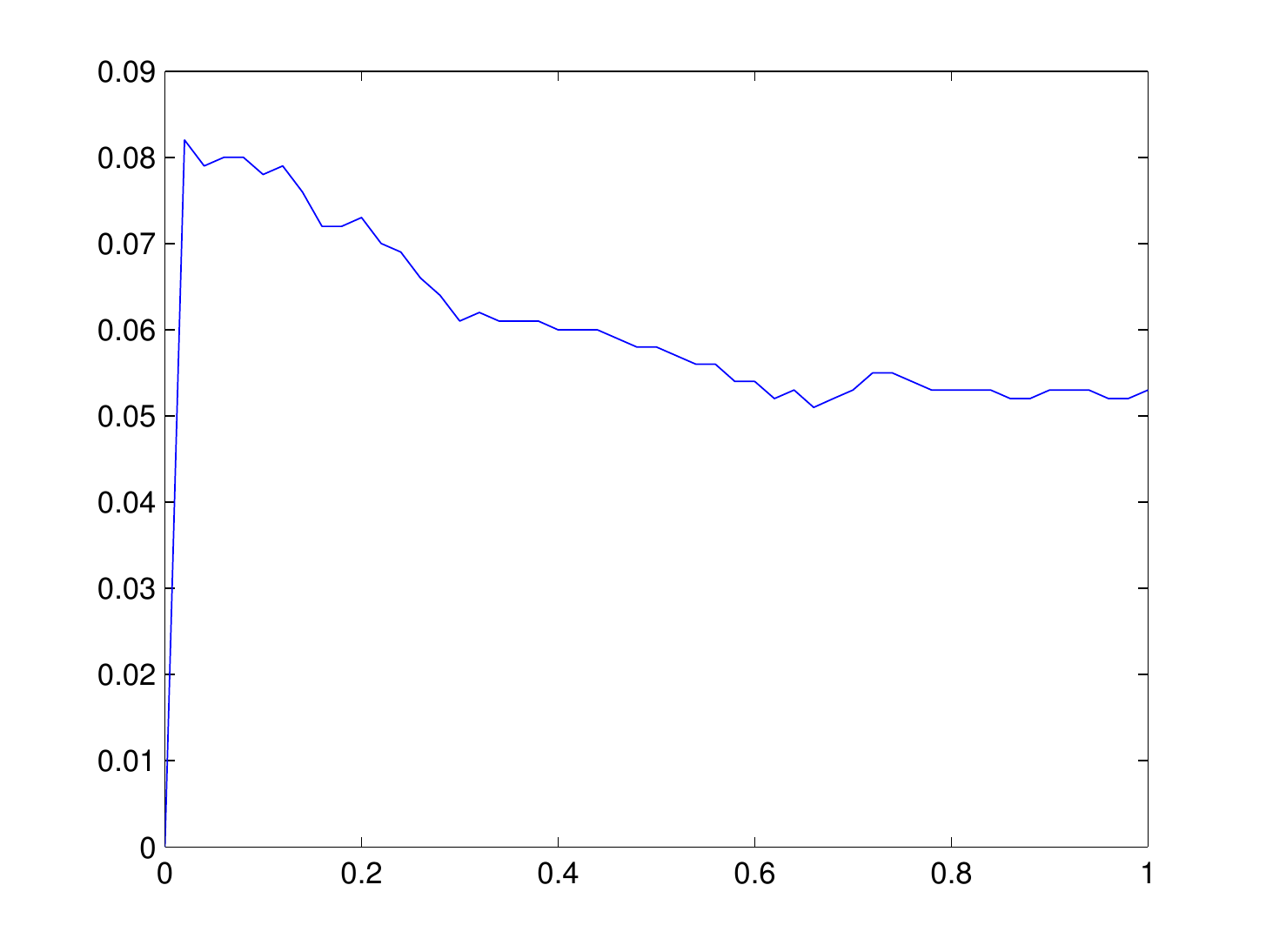}
		\end{tabular}
		\caption{The results of empirical size when covariance functions of two samples are identical (left column) and distinct (right column).
		}\label{fig3}
	\end{center}
\end{figure}

\begin{figure}[H]
	\begin{center}
		\begin{tabular}{cc}
			\includegraphics[height = 3.5cm, width = 4cm]{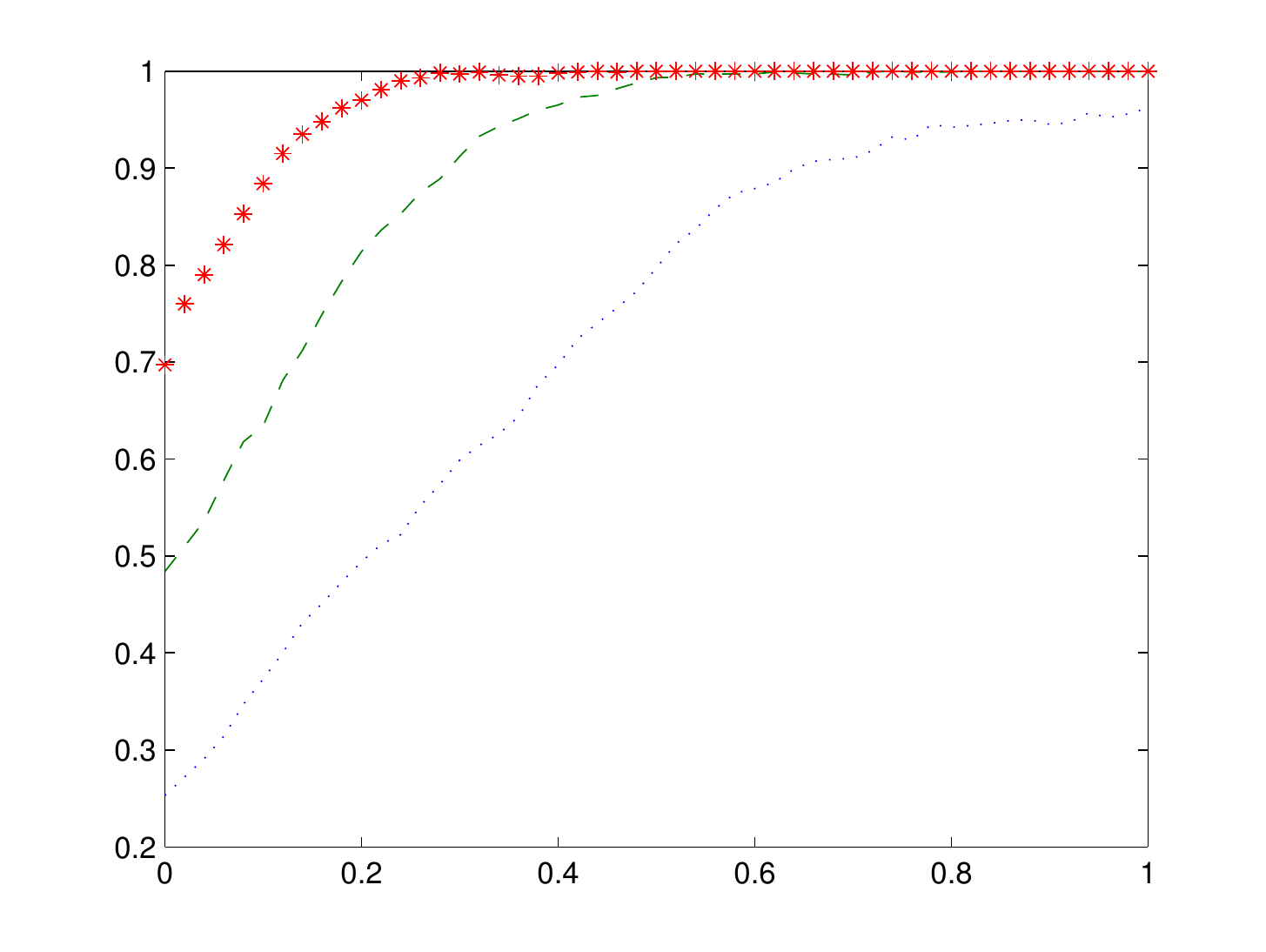}
			\includegraphics[height = 3.5cm, width = 4cm]{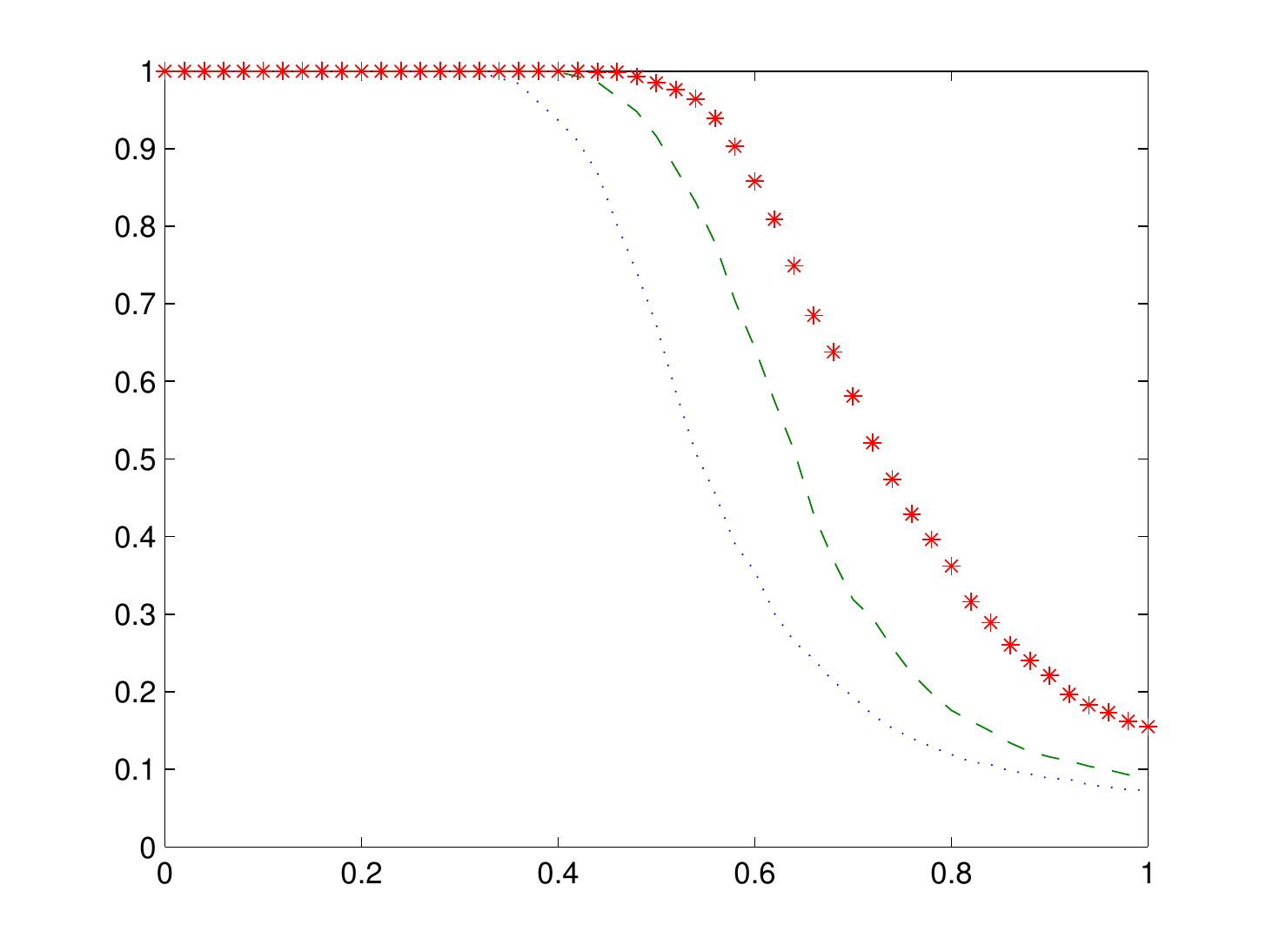}
		\end{tabular}
		\caption{The results of empirical power when covariance functions of two samples are identical (left column) and distinct (right column).
		}\label{fig4}
	\end{center}
\end{figure}

\begin{figure}[h]
	\centering
	\includegraphics[height = 5.5cm, width = 8cm]{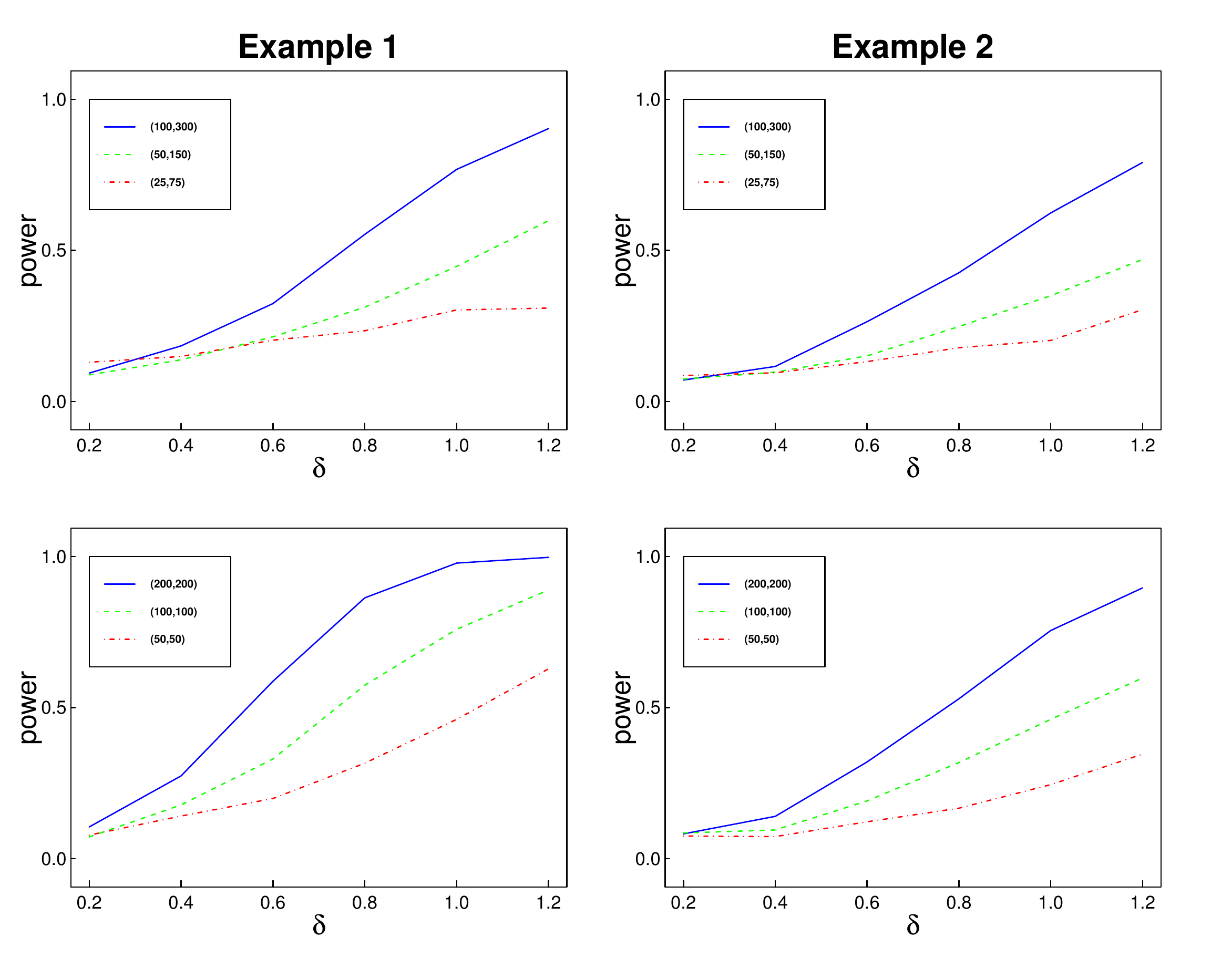}
	\caption{The results of empirical power when covariance functions of two samples are identical (left column) and distinct (right column).
		Top: The results of empirical power  of $(n_1, n_2)= (25,75)$ (red), $ (50,150)$ (green)   and  $(100, 300)$ (blue). Bottom: The results of empirical power  of $(n_1, n_2)= (50,50)$ (red), $ (100,100)$ (green)   and  $(200, 200)$ (blue).
	}\label{fig5}
\end{figure}

Example \ref{eg1} can be seen as two-sample tests where covariance functions are identical, while covariance functions of Example \ref{eg2} are distinct. 
The sample size pair is taken to be $(n_{1},n_{2})=(25,75)$, $(50,150)$, $(100,300)$, $(50,50)$, $(100,100)$, and $(200,200)$, respectively.
The empirical sizes of profile test are computed for different $s$   and $t$.  To save space, we here only present the results of different $s$  for $(n_{1},n_{2})= (100,100)$ in Figure \ref{fig3}. Next, we can also compute the empirical sizes of the globe test. The results are reported in Table \ref{table1}.
The empirical power can be evaluated  when $\delta\neq 0$. The empirical
power at $\delta= 0.4, 0.6, 0.8$  of profile tests are displayed in Figure \ref{fig4} while the results of globe tests at $\delta=0.2, 0.4, 0.6, 0.8, 1.0, 1.2$ are scatter plotted in Figure \ref{fig5}.

\vspace{-1cm}

\begin{table}[H]
	\caption{Empirical sizes  of two proposed test procedures in Examples 1 and 2.}\label{table1}
	\centering\tabcolsep 8pt
	\begin{tabular}{ccccccc}
		\hline
		$(n_{1},n_{2})$&(50,50)&(100,100)&(200,200)&(25,75)&(50,150)&(100,300)  \\
		\hline
		Example 1      & 0.079 &  0.060  &  0.050  & 0.111 & 0.085  & 0.061 \\
		Example 2      & 0.074 &  0.064  &  0.048  & 0.080 & 0.062  & 0.048 \\
		\hline
	\end{tabular}
\end{table}
\vspace{-1cm}

Several observations can be concluded from Figures \ref{fig3} and \ref{fig4}. Firstly, the profile tests have a good control of the type I error. The empirical sizes of identical covariance scenarios are better than that of distinct covariance cases.
Secondly, the empirical power of the test becomes larger  when $\delta$ increases from 0.4 to 0.8,  which is expected. Lastly, the empirical power for the same covariance case is slightly
larger than that of  the  different covariance function cases.


We may observe from Table \ref{table1} and Figure \ref{fig5} that the globe test approach can keep steady empirical size even at pairs of small sample sizes $(n_1, n_2)= (25,75)$ or $(50, 50)$.
The empirical power of two test methods increases as the sample size increases.
When $\delta$ increases from 0.2 to 1.2, the empirical power of the test becomes more and more large, which is evidence of the consistency of the testing procedures. Also the empirical power of equal sample size scenario is slightly better than that of  unequal sample size one.

%

\section{Real data examples}\label{sec6}
To illustrate profile and globe tests methods, we analyse the historical precipitation data in the Midwest of USA  and the period lifetables in Europe for human mortality trend analysis.

\subsection{Precipitation data}
The first example is used to analyze the changes of precipitation during 1941-2000 or in different regions
in the Midwest of USA. \cite{BerkesGabrys2009} detected no changes during the period 1941-2000 for only one station  while \cite{GromenkoKokoszka2017JRSSb} detected the change of precipitation during 1941-2000 over the whole region.

The precipitation data is available from the global historical climatological network database. The comprehensive U.S. Climate Normals dataset includes various derived products including daily air temperature normals, precipitation normals and hourly normals.
The dataset that we analyzed in this paper can be downloaded directly from GHCN (Global Historical Climatology Network)-Daily, an integrated public database of NOAA \url{(https://www.ncdc.noaa.gov/oa/climate/ghcn-daily/)} by an R interface. Our interest is daily precipitation records from Midwestern states  including  Illinois, Indiana, Iowa, Kansas, Michigan, Minnesota, Missouri, Nebraska, North Dakota, Ohio, South Dakota, and Wisconsin.
In Figure \ref{fig1}, totally 59 locations of the climate monitoring stations are indicated with blue circles $\textcolor{blue}{\circ}$ in 4 states from the Great Plains (light green region), and  with red triangles $\textcolor{red}{\triangle}$ in 5 states from the Great Lakes (yellow region). Notice that there is no climate monitoring stations in Iowa, Michigan, and Missouri. 
We target to detect whether the changes of average precipitation took place for different time phases or regions.

Let $Y_{i}(s,t)$ be the precipitation of the $t$th day in the ith year
of the $s$th station.
Before we apply the proposed method, we need to do registration with the data.  To remove the effects due to the heavy tail distribution, we apply
the transformation
\begin{equation*}
	Z_{i}(s,t)=\log_{10}\{Y_{i}(s,t)+1\},
\end{equation*}
where $\{Y_{i}(s,t)\}$ are original records.
After the transformation, we
pre-smooth data by using the cubic splines function. It is noted that  the data of every climate
monitoring stations from 1941 to 2000 can be constituted into a time series with length
21900(365 day by 60 year). Then, the data of the 59 climate monitoring stations can be
seen as a sample with sample size being 21900 and variables being 59. According to the empirical
Pearson correlation of 59 variables, the 59 climate monitoring stations is stringed into a
function
by the stringing method in \cite{ChenChenWangMuller2011JASA}. Consequently the spatiotemporal data $\{Y_{i}(s,t)\}$  are converted into the bivariate functional data  $\{X_{i}(s,t)\}$. Notice that the difference between the spatiotemporal data $Y_{i}(s,t)$ and the bivariate functional data $X_{i}(s,t)$  is that the argument $s$ in the former expression has no order but it is ranked in the latter.

\begin{figure}[H]
	\begin{center}
		\includegraphics[height = 5.5cm, width = 8cm]{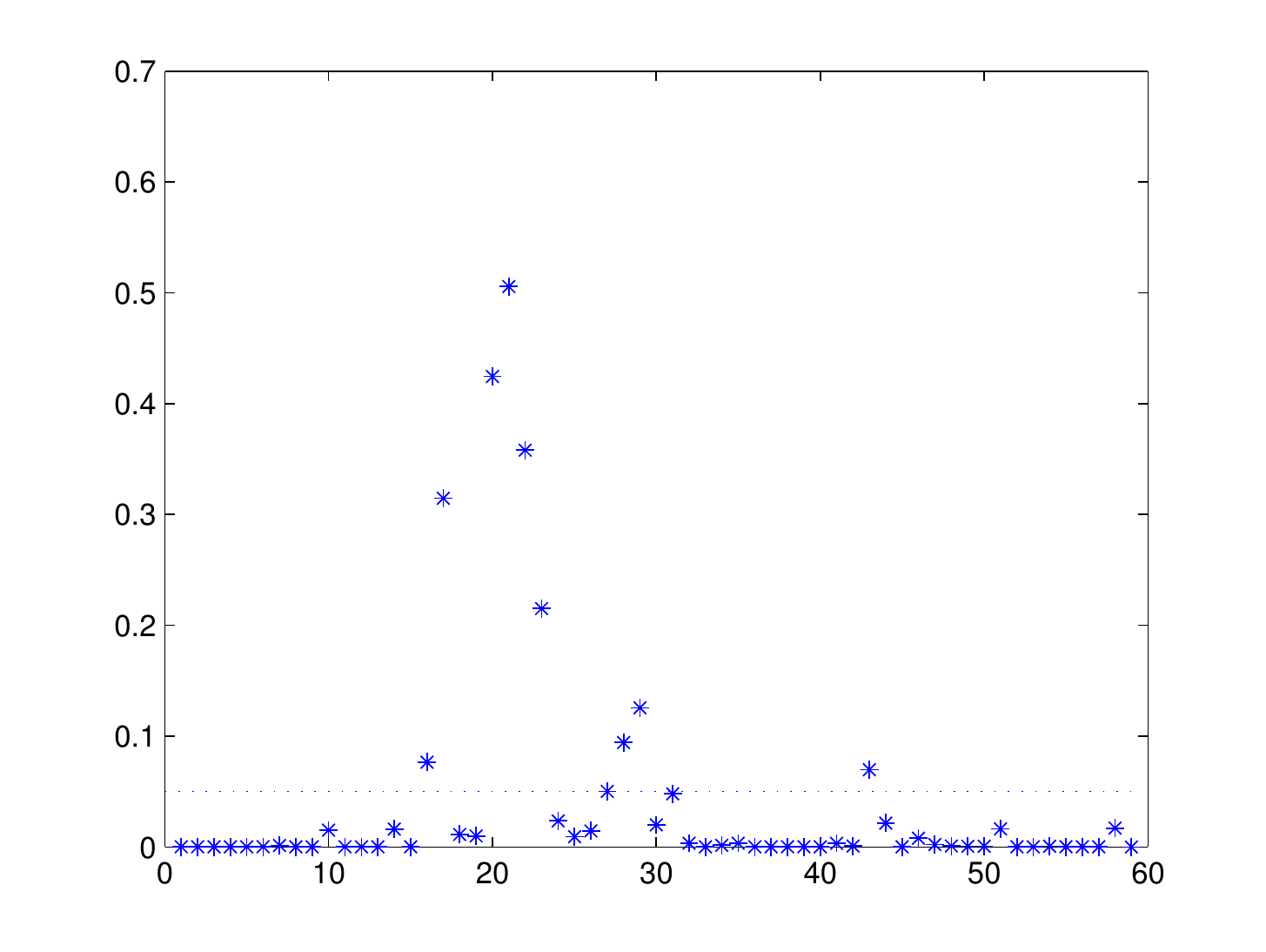}
		\caption{The $p$-value of the profile tests for every station.
		}\label{fig6}
	\end{center}
\end{figure}

\cite{GromenkoKokoszka2017JRSSb} studied the data $Y_{i}(s,t)$ and detected out the change of  the average precipitation at about  1967.   In this subsection,
we firstly apply the profile test to check if the profile of mean surfaces  are equal during the periods 1941-1967 and 1968-2000. It   corresponds to  test  whether  the average precipitation of every
station has changes during these periods. The $p$-values  of  the profile  tests  are computed  and  results are displayed
in Figure \ref{fig6}.  As can be seen from Figure \ref{fig6}, most of the  $p$-values are less than 0.05 or significant except 11  stations.
For ease of reference,  we list  the latitude and longitude
in Table \ref{table2} for 11 stations. This displays that  the average precipitation  of most locations had changed during  the periods 1941-1967 and 1968-2000.
\vspace{-5mm}
\vspace{-5mm}
\begin{table}[H]
	\caption{The latitude and longitude of stations where the $p$-value of profile test are more than 0.1.}
	\centering\label{table2}\tabcolsep 18pt
	\begin{tabular}{ccc}
		\hline\hline
		&&\vspace{-5mm}\\
		Code        & latitude & longitude \\
		\hline
		USC00148235 & 38.4661  & -101.7758\\
		USC00250050 & 42.5522  & -99.8556\\
		USC00252145 & 41.4086  & -102.9661\\
		USC00255090 & 40.8508  & -101.5428\\
		USC00325479 & 46.8128  & -100.9097\\
		USC00394007 & 43.4378  & -103.4739\\
		USC00398307 & 45.4283  & -101.0764\\
		USC00394007 & 43.4378  & -103.4739\\
		USC00392797 & 45.7644  & -99.6353\\
		USC00321871 & 48.9075  & -103.2944\\
		USC00327530 & 46.8886  & -102.3192\\
		\hline\hline
	\end{tabular}
\end{table}
Next,  we implement the following globe test
\begin{eqnarray*}
	\begin{aligned}
		&	H_{0}^{\text{Midwest}}: \mu_{1}(s,t)=\mu_{2}(s,t)\\
		&	\text{vs.} \hspace{4pt} H_{1}^{\text{Midwest}}: \mu_{1}(s,t)\neq \mu_{2}(s,t), s\in \mathbb{R}^{59}, t \in \mathbb{R}^{365}.
	\end{aligned}
\end{eqnarray*}
From the globe  test procedure presented in Section \ref{sec4} together with the asymptotic distribution
of the test statistic   $\widehat{\text{TM}}$, we calculate the corresponding $p$-value to be  0.001.  This result is consistent  with the conclusion of Gromenko, Kokoszka and Reimherr. That is, the patterns of mean surfaces
are different over the whole Midwest region between  before 1967 and after 1967.  Intuitively,
according to the results of the profile test, the precipitation had changed  in  most of locations  which lead to  the variations  of whole region.


The heatmaps in Figure \ref{heatmap} leak the information that sample mean values of annual precipitation in the Great Lakes (GL) based on 28 stations are more than that in the Great Plains (GP). This motivates us to further explore how the mean functions of bivariate functional data $\{X_{i}(s,t)\}$ was affected  by temporal and spatial effects from both domains. It is natural to test the equality of two mean surfaces of the precipitation for the 31 stations located in the GP and the 28 stations located in the GL during the periods 1941-1967 and 1968-2000, respectively by
\[ H_{0}^{1967-}: \mu^{\text{GP}}=\mu^{\text{GL}}\hspace{4pt}
\text{vs.} \hspace{4pt}
H_{1}^{1967-}: \mu^{\text{GP}}\neq\mu^{\text{GL}}, \]
and
\[ H_{0}^{1967+}: \mu^{\text{GP}}=\mu^{\text{GL}}\hspace{4pt}
\text{vs.} \hspace{4pt}
H_{1}^{1967+}: \mu^{\text{GP}}\neq\mu^{\text{GL}}. \]
All the $p$-values by globe test procedures for above two hypotheses are tiny approaching to zero indicating rejecting the null hypotheses but in favor of the alternative one. It is consistent with the intuition that the mean patterns of precipitation at Great Plains and at Great Lakes are different.

Furthermore,
for the 28 stations located in the GL, we test the mean surfaces of precipitation before and after 1967,  denoted by
\[ H_{0}^{\text{GL}}: \mu^{1967-}=\mu^{1967+}\hspace{4pt}
\text{vs.} \hspace{4pt}
H_{1}^{\text{GL}}: \mu^{1967-}\neq\mu^{1967+}. \]
The $p$-value is 0.0163. The null hypothesis would be rejected at 0.05 significance level.
Testing equality of the mean surfaces of precipitation before and after   1967  is also implemented for the 31 stations located in the GP,
denoted  by
\[ H_{0 }^{\text{GP}}: \mu^{1967-}=\mu^{1967+}\hspace{4pt}
\text{vs.} \hspace{4pt}
H_{1}^{\text{GP}}: \mu^{1967-}\neq\mu^{1967+}. \]
The $p$-values by globe testing method are 0.5677. The null hypothesis would not be rejected at 0.05 significance level.
That is, averagely speaking, the precipitation in the Great Lakes changed before 1967 and after 1967, whereas the mean pattern of precipitation in the Great Plains had no change before 1967 and after 1967. \emph{Therefore, our analysis provides evidence that change in the mean function of precipitation was mainly due to the Great Lakes but the Great Plains may be affected little.  By looking up the map,  we find that all the stations in Table \ref{table2}  are located in the Great Plains. It further verify the reliability of the proposed methods.} All testing results are presented in Table \ref{table3}.


\begin{table}[H]
	\caption{\label{table3} Results of the tests  based on statistics  $\widehat{\text{TM}}$.}\label{ms-1} \centering\tabcolsep 15pt
	\begin{tabular}{ccc}
		\hline\hline
		&&\vspace{-8mm}\\
		statistic    & \text{the observed value of a statistic}& $p$-value  \\
		\hline\vspace{-8mm}\\
		\multicolumn{3}{c}{$H_{0}^{\text{Midwest}}: \mu^{1967-}(s,t)=\mu^{1967+}(s,t)$}\\
		\hline\vspace{-8mm}\\		
		$\begin{array}{c}
		\widehat{\text{TM}}
		\end{array}$
		& $\begin{array}{c}
		123.6
		\end{array}$
		&$\begin{array}{c}
		\textbf{0.001}
		\end{array}$ \\
		\hline \vspace{-8mm}\\		
		\multicolumn{3}{c}{$H_{0}^{\text{GP}}: \mu^{1967-}(s,t)=\mu^{1967+}(s,t)$}\\
		\hline\vspace{-8mm}\\		
		$\begin{array}{c}
		\widehat{\text{TM}}
		\end{array}$
		& $\begin{array}{c}
		59.7175
		\end{array}$
		&$\begin{array}{c}
		0.5677
		\end{array}$ \\		
		\hline \vspace{-8mm}\\
		\multicolumn{3}{c}{$ H_{0}^{\text{GL}}:\mu^{1967-}(s,t)=\mu^{1967+}(s,t)$}\\
		\hline\vspace{-8mm}\\		
		$\begin{array}{c}
		\widehat{\text{TM}}
		\end{array}$
		& $\begin{array}{c}
		108.20
		\end{array}$
		&$\begin{array}{c}
		\textbf{0.0163}
		\end{array}$ \\
		\hline\vspace{-8mm}\\		
		\multicolumn{3}{c}{$H_{0}^{1967-}:\mu^{\text{GP}}(s,t)=\mu^{\text{GL}}(s,t)$}\\
		\hline\vspace{-8mm}\\		
		$\begin{array}{c}
		\widehat{\text{TM}}
		\end{array}$
		& $\begin{array}{c}
		973.11
		\end{array}$
		&$\begin{array}{c}
		\textbf{0.0000}
		\end{array}$ \\
		\hline\vspace{-8mm}\\		
		\multicolumn{3}{c}{$ H_{0}^{1967+}:\mu^{\text{GP}}(s,t)=\mu^{\text{GL}}(s,t)$}\\
		\hline\vspace{-8mm}\\		
		$\begin{array}{c}
		\widehat{\text{TM}}
		\end{array}$
		& $\begin{array}{c}
		1116.4
		\end{array}$
		&$\begin{array}{c}
		\textbf{0.0000}
		\end{array}$ \\
		\hline		
	\end{tabular}
\end{table}
\vspace{-5mm}
\subsection{European human mortality rate data}
In the second example, we will analyse the trends in human mortality based on the records in the period life tables during the calendar years 1960-2006 for Europe countries. A period life table represents the mortality conditions at a specific moment in time. It is approachable from the Human Mortality Database via the website linkage \url{www.mortality.org} (\citealp{WilmothAndreev2007}). The analysis of trends in human mortality is important to recover  the demographic impacts.  Results of such research will benefit the prediction and forecasting of future cohort mortality \citep{Vaupel1998, OeppenVaupel2002}. We focus on  comparison of different countries or genders,   specifically   on the  older ages over 50 years old.

\begin{figure}[H]
	\begin{center}
		\begin{tabular}{cc}
			&\hspace{-7mm}\includegraphics[height = 3cm, width = 4cm]{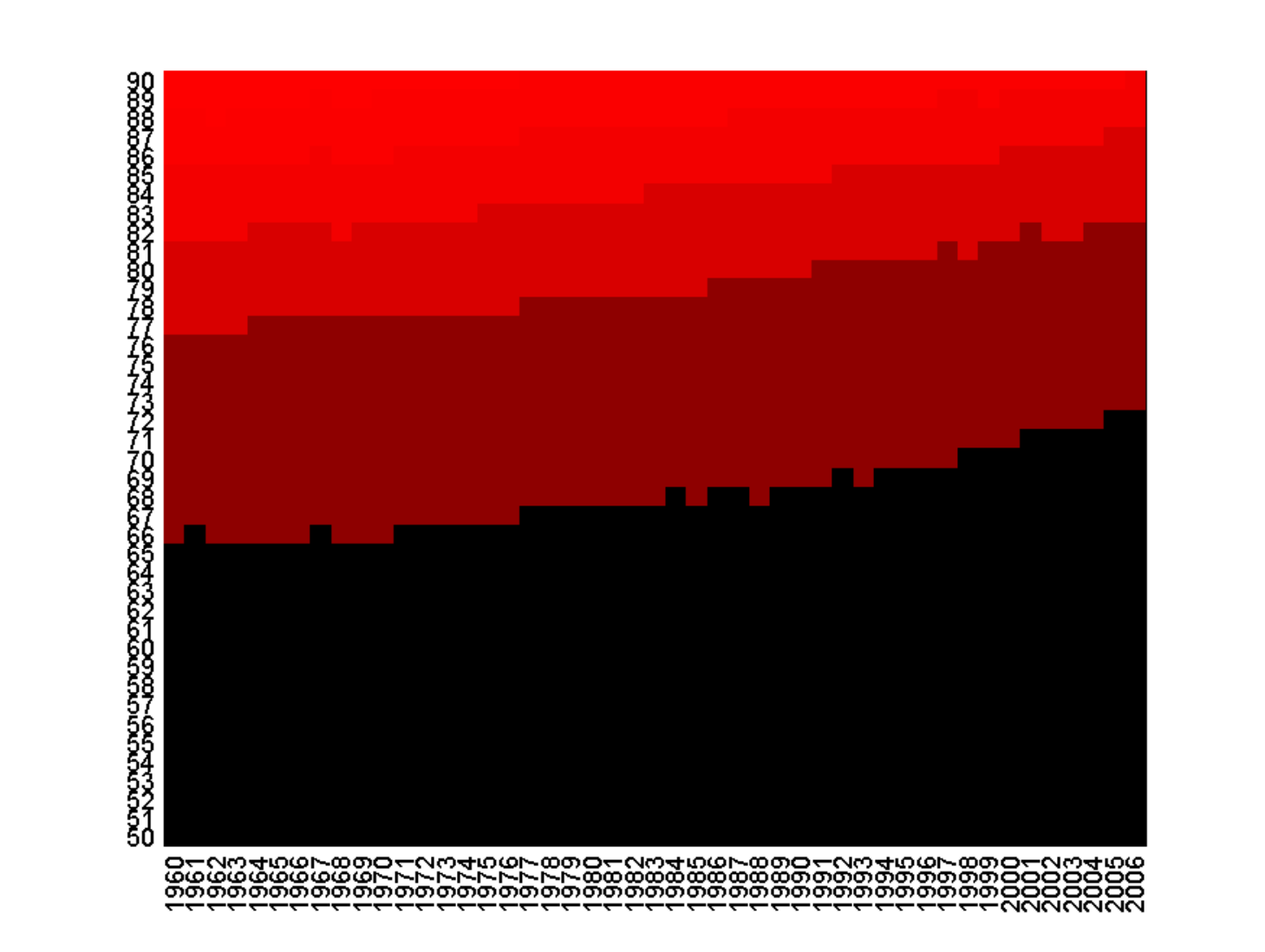}	
			\includegraphics[height = 3cm, width = 4cm]{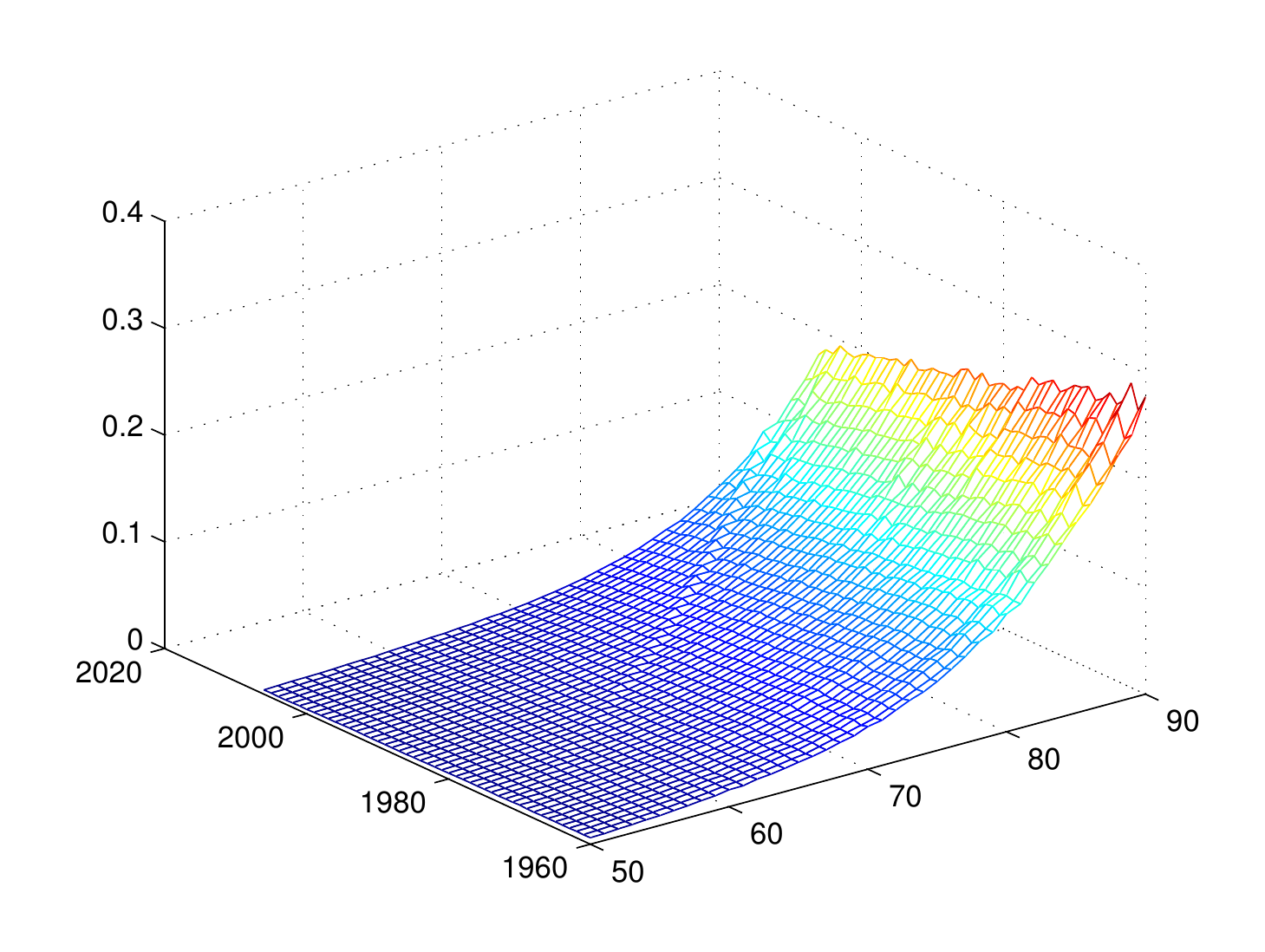}\\
			&\hspace{-7mm}\includegraphics[height = 3cm, width = 4cm]{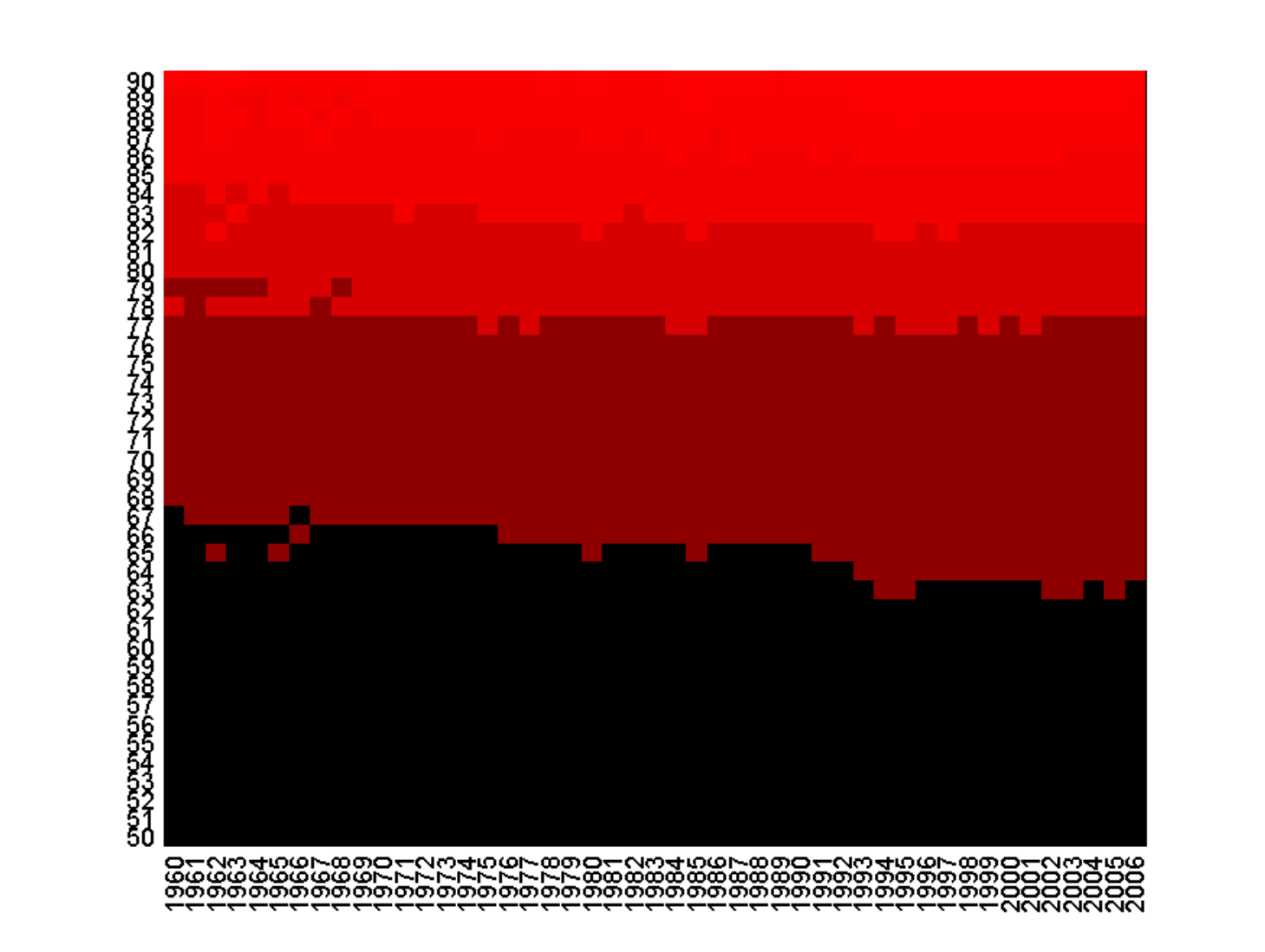}
			\includegraphics[height = 3cm, width = 4cm]{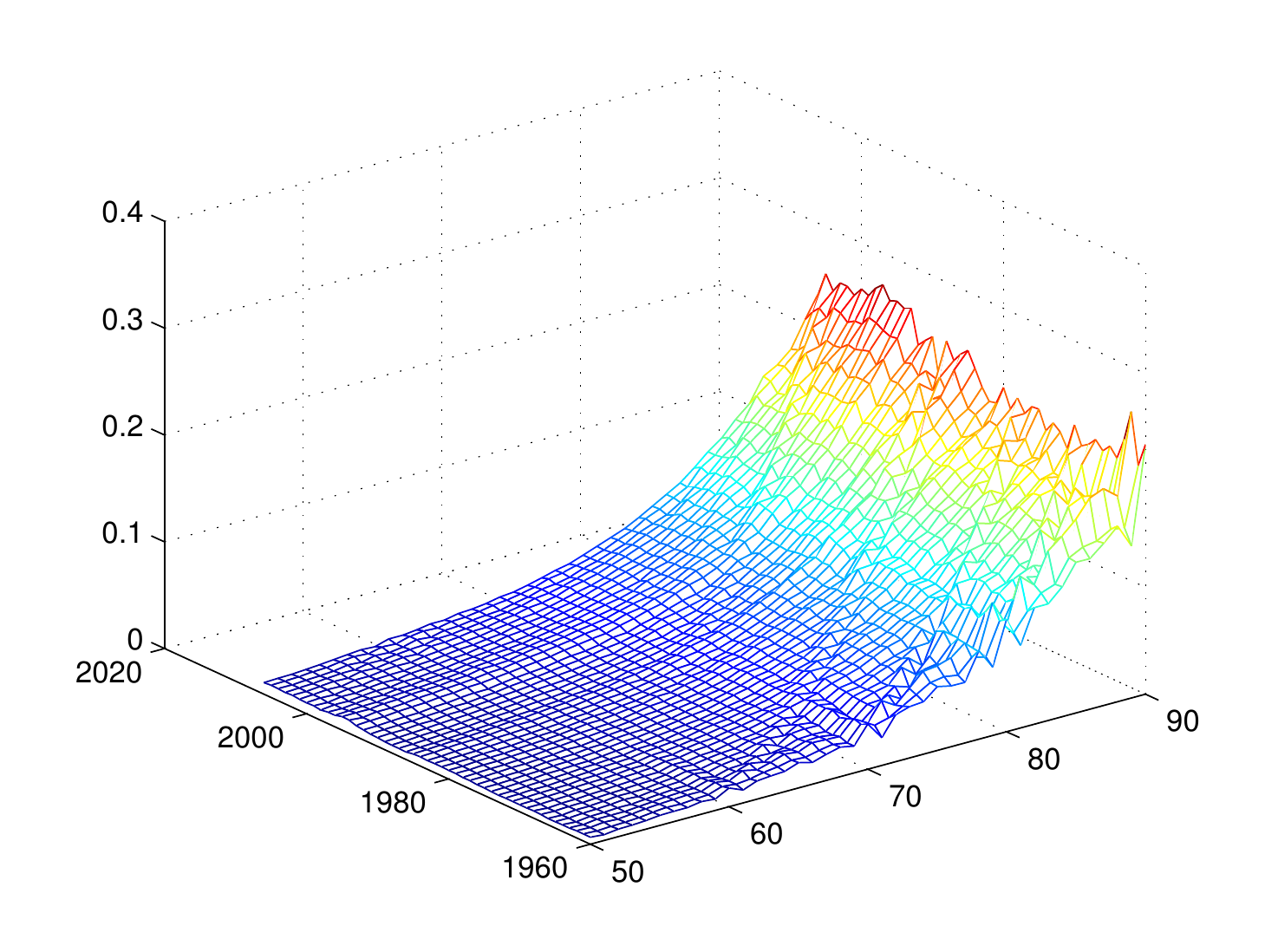}\\
		\end{tabular}
		\caption{Top:  Sample means of  the mortality
			rate for the 5 Eastern European countries. Bottom: Sample means of    the mortality rate for the 27 Western European
			countries.}\label{fig7}
	\end{center}
\end{figure}

\begin{figure}[H]
	\begin{center}
		\begin{tabular}{cc}
			\hspace{-5mm}
			\includegraphics[height = 3.3cm, width = 4cm]{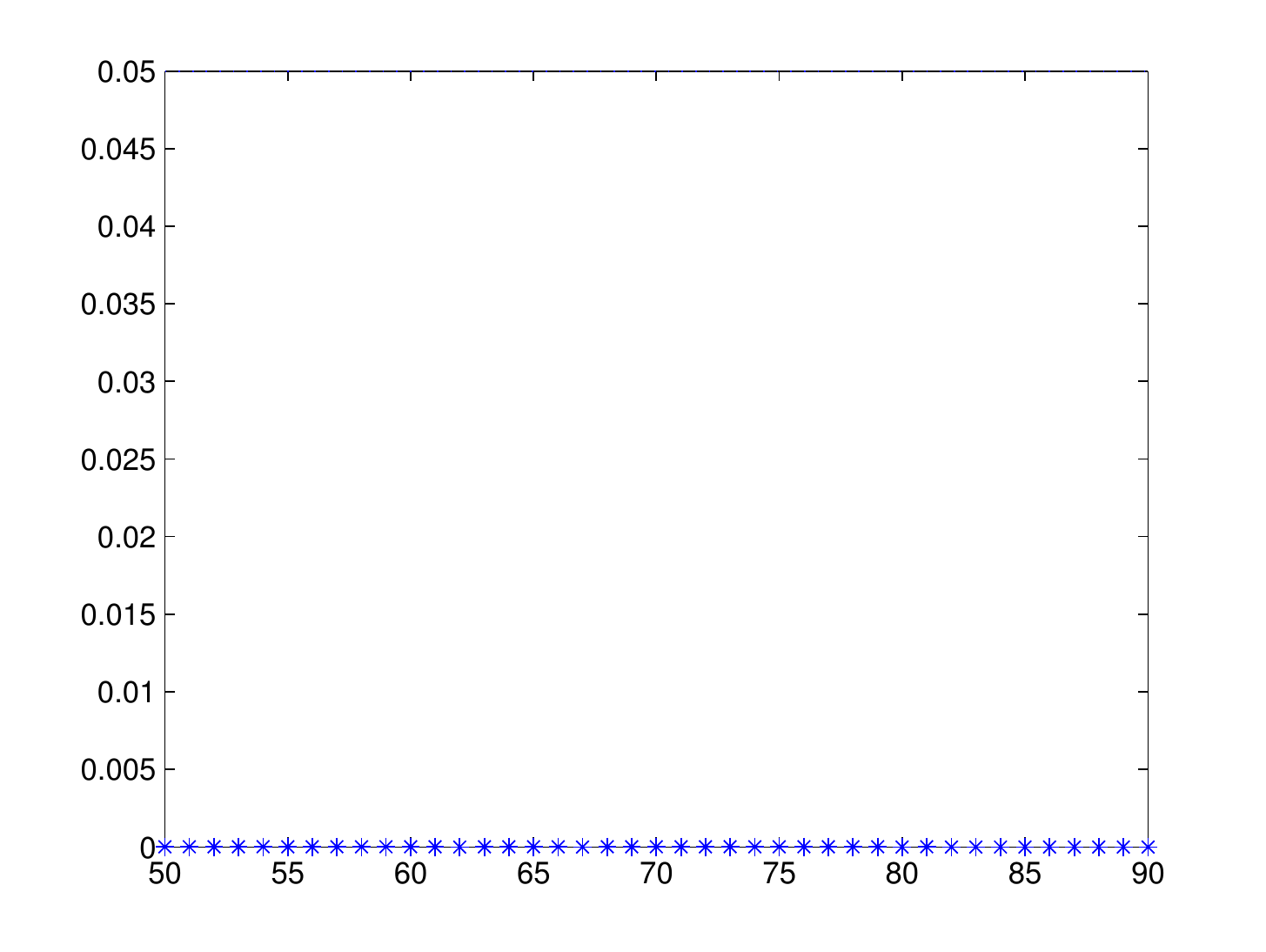}
		    \includegraphics[height = 3.3cm, width = 4cm]{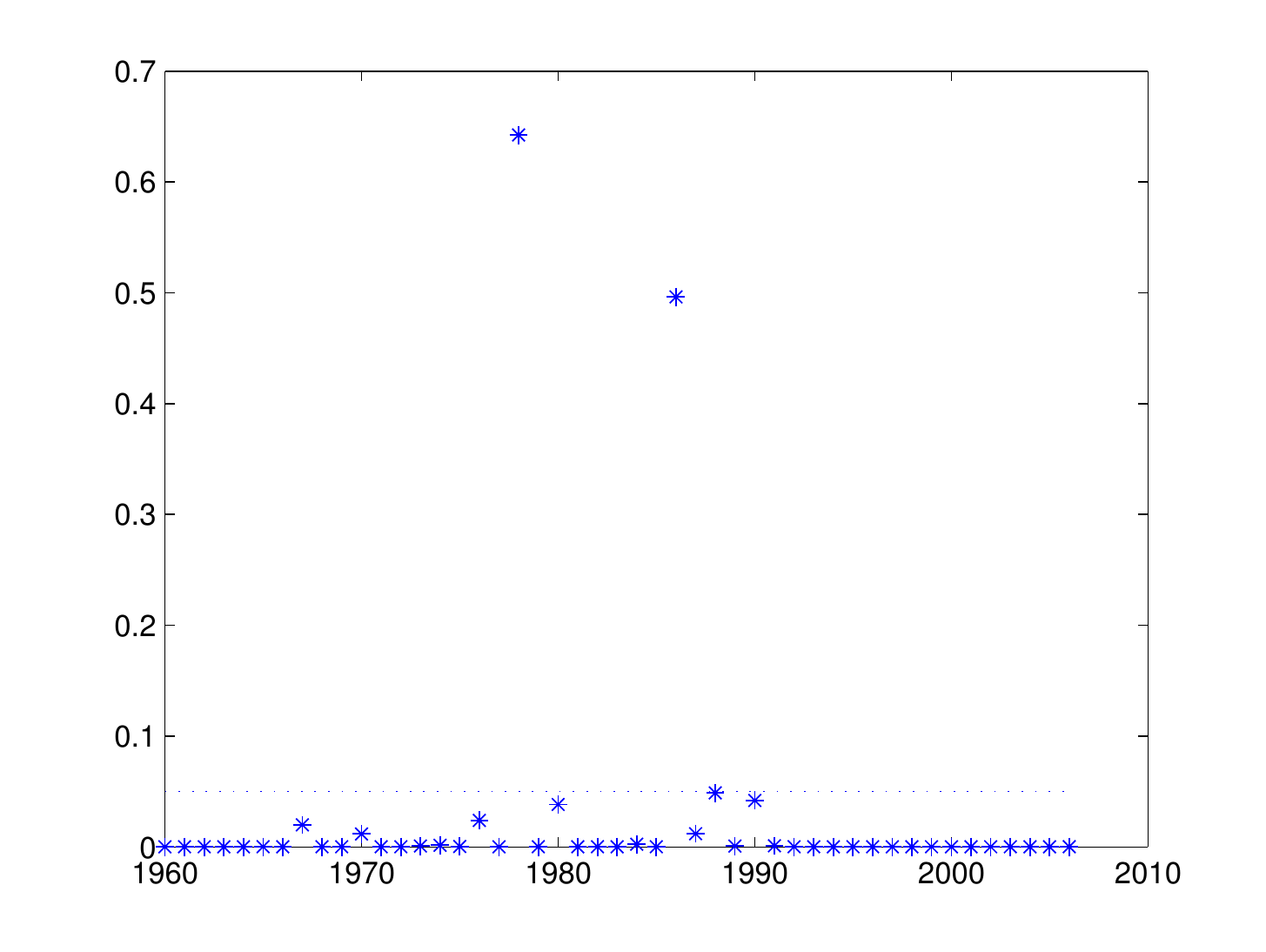}
		\end{tabular}
		\caption{The $p$-value  of  the profile  tests   for every age (left) and year(right).
		}\label{fig8}
	\end{center}
\end{figure}

There are 32 countries included in the European period life tables. It contains five Eastern European countries, Belarus, Bulgaria, Russia, Ukraine and Lithuania, and the remaining 27 Western European countries.
Following the notation introduced in Section \ref{sec3},
$X_{i}^{(1)}(s, t), i=1,\ldots,5, $  denotes the mortality rate of
the  five  Eastern European countries for subjects at age $s$ and calendar year $t$,  where $50 \leq s \leq 90$, focusing on
the death rates of older individuals, and on a recent block of 47 years, $1960 \leq t \leq 2006$.  Similarly,  $X_{i}^{(2)}(s, t), i=1,\ldots,27, $  denotes the mortality rate for other countries.
The sample  mean function  $\widehat{\mu}_{1}(s, t)=\sum_{i=1}^{5}X_{i}^{(1)}(s, t)$  and  $\widehat{\mu}_{2}(s, t)=\sum_{i=1}^{27}X_{i}^{(2)}(s, t)$  for two clusters of countries are visualized in Figure \ref{fig7}.
The heatmaps and sample mean surfaces show obvious opposite trend of mortality rates particularly for very aged people in Eastern and Western European countries as the calendar year passed 1980 or so. We apply the profile and globe test procedures to test  if the two underlying mean surfaces and its profile are different.
\vspace{-5mm}
According to profile test method introduced in Section \ref{sec3},  we implement the tests \eqref{2.1a} and \eqref{2.1b}.
The $p$-values
for fixed $s^{*}$ or $t^{*}$ are calculated, respectively. The results are presented in Figure \ref{fig8}.
For every fixed age $s^*$, we find that all of $p$-values are approaching to zero. This indicates that the mean mortality rates of the Eastern and Western European
is different for every age $s^{*}= 50,\cdots,90$. For every fixed year $t^*$, almost all $p$-values are less than 0.05
except for years $t^{*}=1978$ and $ 1986$.
Sequentially, we implement the globe test for the mean mortality rates of the Eastern and Western European.
The numbers of included components is $J = 2, K_1=2, K_2=2$  are  chosen by the fraction of variance explained (FVE) criterion with the threshold 0.90.
Based on the asymptotic distribution of the test statistic   $\widehat{\text{TM}}$, the $p$-value is calculated to be 0. It coincides with the intuition on images in Figure \ref{fig7} and is evidence that the mean surfaces of the mortality rates  are different between  the Eastern and Western European countries. Also, it is consistent with  the conclusion of the profile test because almost of
$H_{0}^{\mathcal{S}}$  and $H_{0}^{\mathcal{T}}$  are rejected for fixed  $s^*$ and $t^*$.

\vspace{-5mm}

\begin{figure}[H]
	\begin{center}
		\begin{tabular}{cc}
			&\hspace{-7mm}\includegraphics[height = 3cm, width = 4cm]{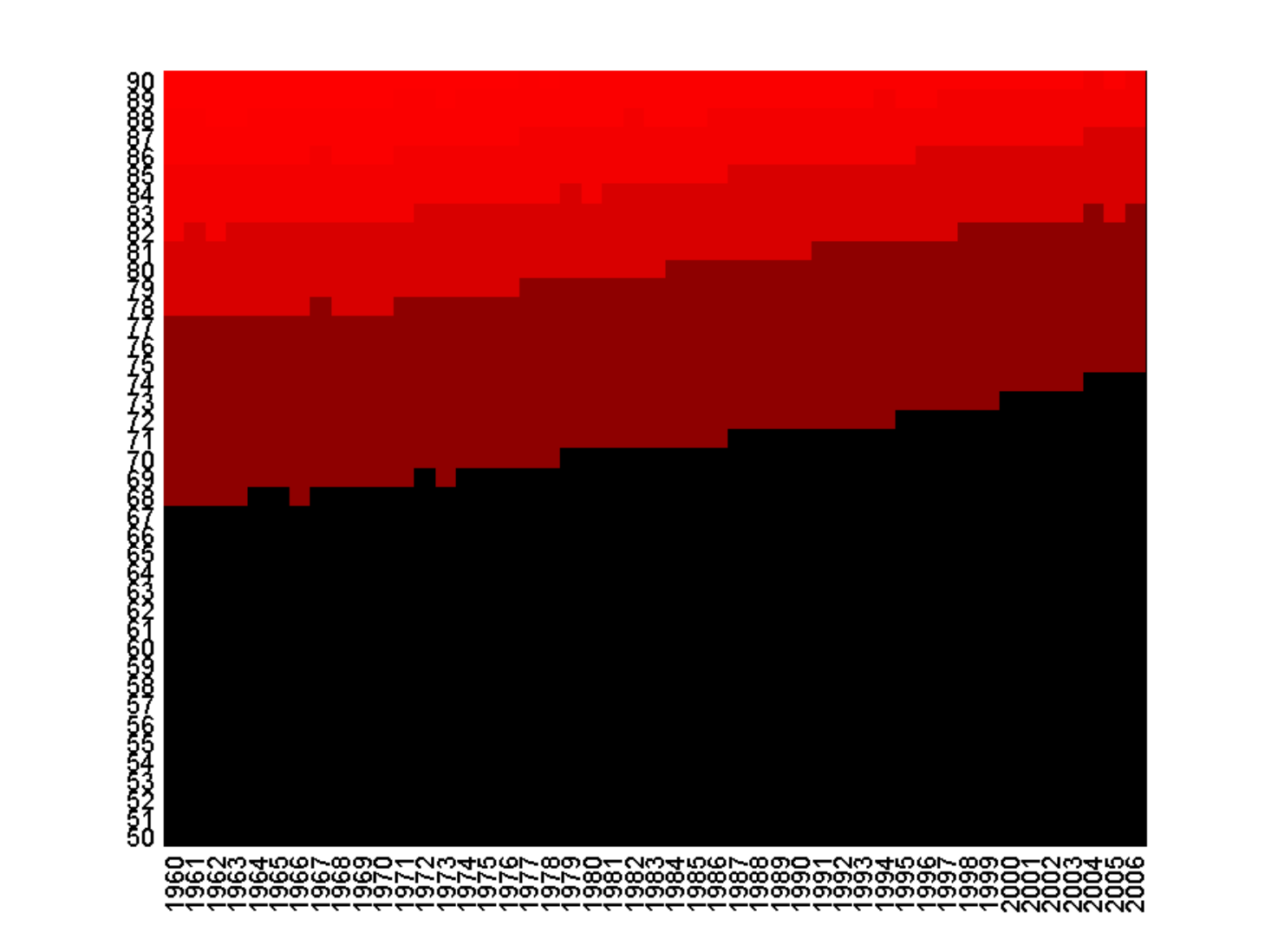}	
              \includegraphics[height = 3cm, width = 4cm]{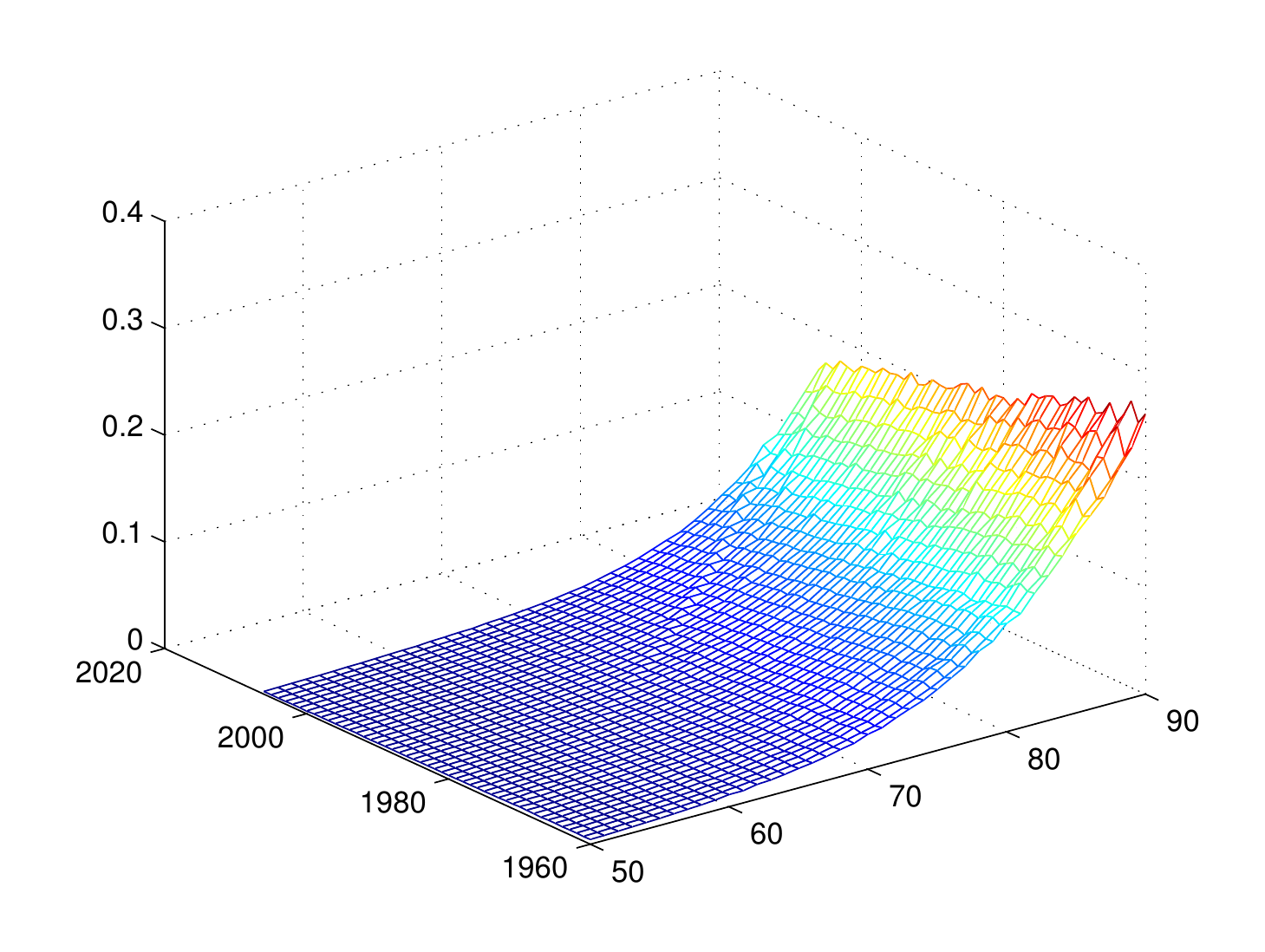}\\
            &\hspace{-7mm}\includegraphics[height = 3cm, width = 4cm]{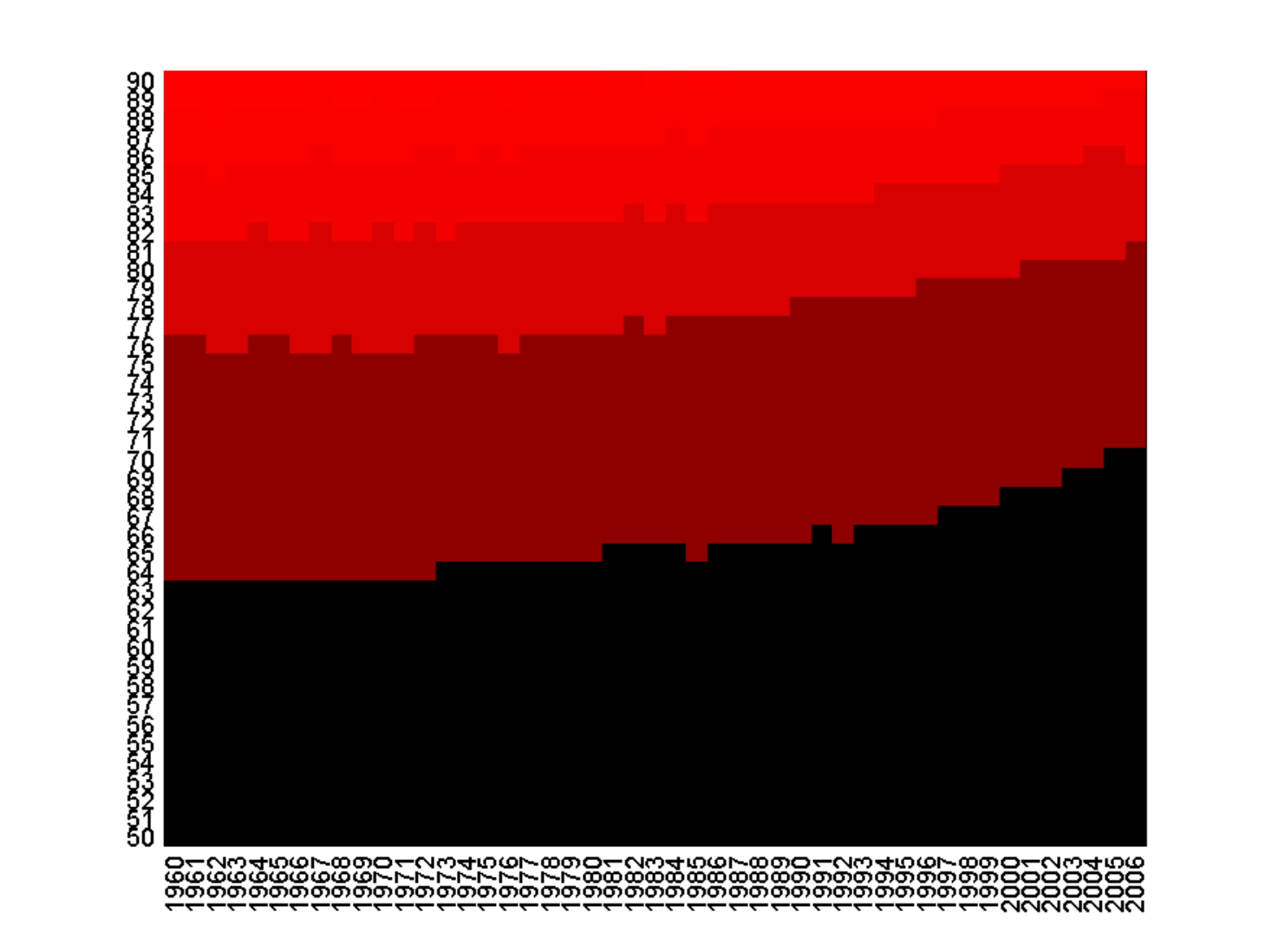}
              \includegraphics[height = 3cm, width = 4cm]{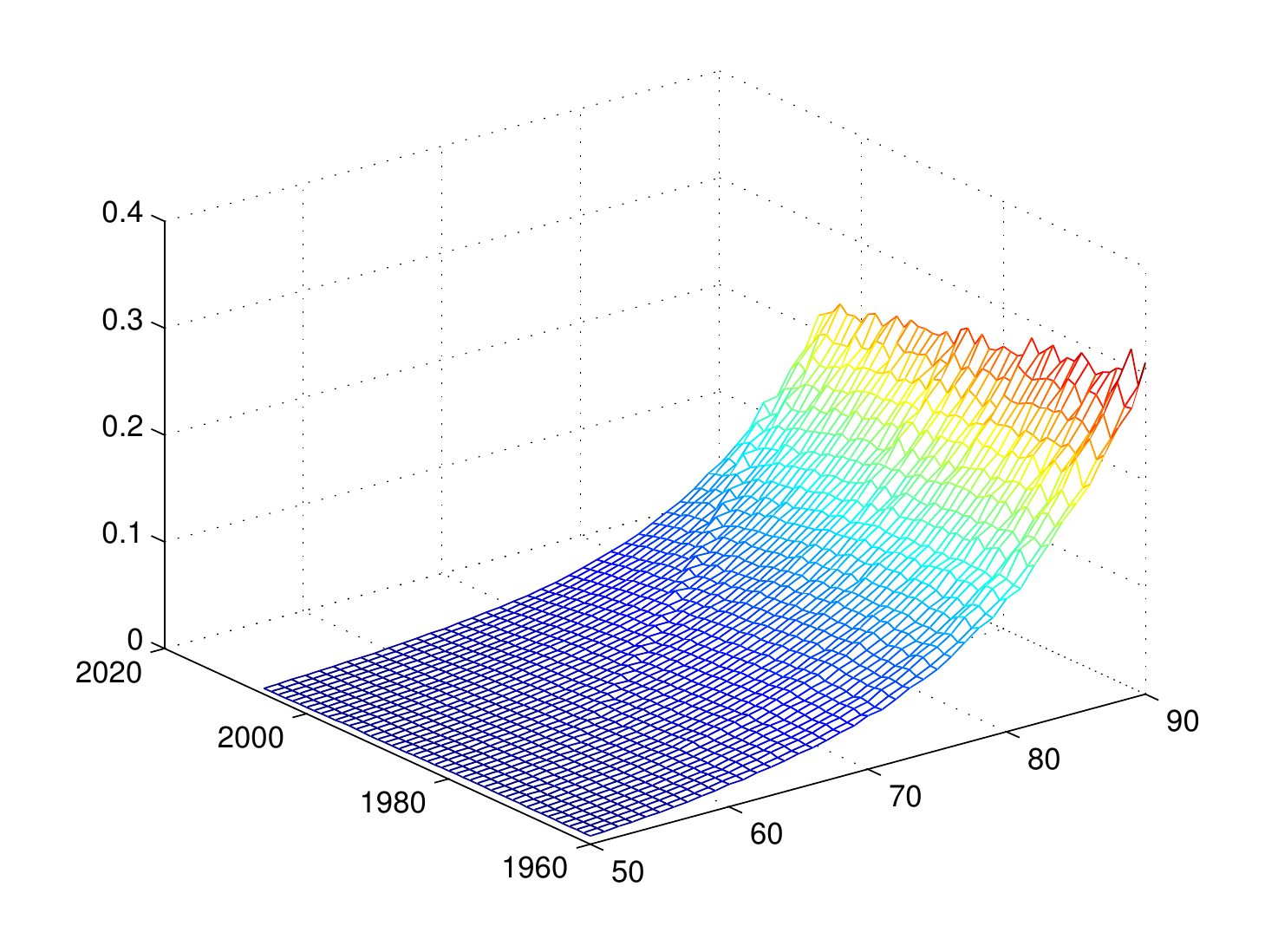}\\
		\end{tabular}
		\caption{Top: Sample means of the mortality rate of male. Bottom: Sample means of the mortality rate of female.}\label{fig9}
	\end{center}
\end{figure}

Next we examine the equality of mean surfaces and its profile  between female and male clusters in West Europe.
The heatmaps and sample mean surfaces for male and female clusters    are displayed in Figure \ref{fig9}. Intuitively it does not show obvious difference. However,
all the $p$-values of profile tests are zero for fixed $s^*$ and $t^*$.
Furthermore, we also implement globe test and obtain the $p$-value that is 0. Therefore, the mean surface and its profile are different in Western Europe for aged people in different gender type.

\vspace{-7mm}

\section{Discussion}\label{sec7}
Bivariate functional data have been definitely presented  in \cite{ParkStaicu2015}, \cite{ChenDelicadoMuller2017} and\\ \cite{AstonPigoliTavakoli2017}.  However, inferential in testing procedures of such data type has not been adequately dealt with in the literature.
This paper develops profile and globe  tests to detect if    mean surfaces and their profile are different for  two bivariate functional samples.

In this paper, for simplicity, we assume $X^{(m)}(s,t)$ are recorded on a regular and dense grid of pair time point. It is noted that the mean function and its profile estimation in \eqref{3.3} and \eqref{4.3} can always be obtained in the sparse and irregular setting  by additional smoothing steps.
Our proposed profile and globe tests can also be implemented if marginal eigenfunctions  and mixed  eigenfunctions can be effectively estimated. Hence, the proposed testing methodology will have wider application and much
more flexible framework.

Additionally, for  bivariate functional data $X^{(m)}(s,t)$,   product  functional principal component analysis (PFPCA) due to \cite{ChenDelicadoMuller2017} and double functional principal component analysis (DFPCA) due to \cite{ChenMuller2012} have been developed. Using the methodology similar to Section \ref{sec3} and \ref{sec4},  profile and globe tests  based on PFPCA and DFPCA can also be developed.

Higher dimensional functional data also occur in practice. For example, in a recent environment and conservation project, the raw water sample has been collected weekly from different branch streams of Dongjiang at Pearl River Delta, China. Quite a few  water quality indices were measured for each sample.  The  measurements database  form
naturally a trivariate functional data.  To save cost and to monitor water quality more effectively, we are interested in detecting whether water quality has significantly changed across time, locations and water quality indices. The corresponding null hypothesis is thus then
\begin{eqnarray*}
	H_{0}:  \mu_1(b,s,t)=\mu_2(b,s,t)  \hspace{4pt}  \text{vs.}  \hspace{4pt} H_{1}: \hspace{4pt} \mu_1(b,s,t)
	\neq \mu_2(b,s,t),
\end{eqnarray*}
where $ \mu_i(b,s,t)$ is the mean function for the $i$th sample $\{X_{i}(b,s,t)\}_{i=1}^{n_i}, i=1, 2$ with sample size $n_i$  collected on time $t$ at location $s$ with water quality index $b$ measured. Although as mentioned earlier that the methods developed can be similarly extended to more than two samples, technical derivations become tedious.
We are currently working on methods for trivariate or higher order multivariate functional data.

\bigskip
\begin{center}
{\large\bf SUPPLEMENTARY MATERIAL}
\end{center}

\begin{description}

\item[Proof of Theorem]
This file is to present the detail of the proof procedure of the corresponding theorems in the article. (file type: pdf)

%

\end{description}

\bibliographystyle{ECA_jasa}
\bibliography{Ref_JBES_Feb24}
\end{document}